\def\year{2019}\relax
\documentclass[letterpaper]{article} 

\usepackage{aaai20}  

\usepackage{times}  
\usepackage{helvet}  
\usepackage{courier}  
\usepackage{url}  
\usepackage{graphicx}  

\usepackage[dvipsnames]{xcolor}
\usepackage{multirow}
\usepackage{colortbl}
\usepackage{tabularx}
\usepackage{booktabs}
\usepackage{collcell}
\usepackage{longtable}
\usepackage{cleveref} 
\usepackage{verbatim} 
\usepackage{amsmath}  
\usepackage{subcaption}
\usepackage{bbding} 
\usepackage{algpseudocode} 
\usepackage{algorithm}
\usepackage{csquotes}
\usepackage{eurosym}
\usepackage{etoolbox,siunitx}
\robustify\bfseries
    \DeclareSIUnit\eur{\officialeuro}
    \DeclareSIUnit\M{M}
    \DeclareSIUnit\k{k}

\newcommand{\mycite}[1]{\citeauthor{#1} (\citeyear{#1})} 
\newcommand{\egcite}[1]{(\eg, \citeauthor{#1} \citeyear{#1})}
\newcommand{\egcitetwo}[2]{(\eg, \citeauthor{#1} \citeyear{#1}; \citeauthor{#2} \citeyear{#2})}
\algnewcommand{\LineComment}[1]{\State \(\triangleright\) #1}

\usepackage{xspace}
	\newcommand\ie{i.\,e.\xspace}
	\newcommand\eg{e.\,g.\xspace}

\usepackage{xcolor}

\usepackage{colortbl}

\nocopyright

\begin{document}

\title{Leveraging Mobility Flows from Location Technology Platforms to \\Test Crime Pattern Theory in Large Cities}

\author{Cristina Kadar\thanks{Corresponding author.}\thanks{Most of this work was carried out while Cristina Kadar was a visitor at the University of Cambridge.}\\ ETH Zurich\\ ckadar@ethz.ch\\ 
\And Stefan Feuerriegel\\ ETH Zurich\\ sfeuerriegel@ethz.ch\\ 
\And Anastasios Noulas\\ New York University\\ noulas@nyu.edu\\ 
\And Cecilia Mascolo\\ University of Cambridge\\ cm542@cam.ac.uk\\}

\maketitle

\begin{abstract}
Crime has been previously explained by social characteristics of the residential population and, as stipulated by crime pattern theory, might also be linked to human movements of non-residential visitors. Yet a full empirical validation of the latter is lacking. The prime reason is that prior studies are limited to aggregated statistics of human visitors rather than mobility flows and, because of that, neglect the temporal dynamics of individual human movements. As a remedy, we provide the first work which studies the ability of granular human mobility in describing and predicting crime concentrations at an hourly scale. For this purpose, we propose the use of data from location technology platforms. This type of data allows us to trace individual transitions and, therefore, we succeed in distinguishing different mobility flows that (i)~are incoming or outgoing from a neighborhood, (ii)~remain within it, or (iii)~refer to transitions where people only pass through the neighborhood. Our evaluation infers mobility flows by leveraging an anonymized dataset from Foursquare that includes almost 14.8~million consecutive check-ins in three major U.S. cities. According to our empirical results, mobility flows are significantly and positively linked to crime. These findings advance our theoretical understanding, as they provide confirmatory evidence for crime pattern theory. Furthermore, our novel use of digital location services data proves to be an effective tool for crime forecasting. It also offers unprecedented granularity when studying the connection between human mobility and crime.  
\end{abstract}


\section{Introduction}


Crime is pervasive in everyday life. In the US alone, the total number of reported crimes amounts to 2,837 per 100,000 inhabitants.\footnote{\tiny\url{ucr.fbi.gov/crime-in-the-u.s/2016/crime-in-the-u.s.-2016/topic-pages/tables/table-1}} This has considerable negative effects for individuals and society as a whole, including financial losses, physical harm, psychological distress, and a reduced quality of life \cite{Doran2012}. Consequently, it is of utmost importance for research to advance our understanding of crime and to support public decision-makers by developing strategies for effective crime prevention.


Theoretical works in criminology research have conceptualized the spatio-temporal characteristics of crime as the primary entity of interest \egcite{Johnson2010}. It is well established that crime does not occur randomly or uniformly in space and time. However, most theoretical formalizations neglect the fact that humans are also subject to spatio-temporal dynamics. Rather, crime is the result of complex interactions between motivated offenders and unguarded victims/targets and, therefore, it occurs when these elements presumably converge \cite{Cohen1979}. 

\begin{figure}[htb]
\centering
\includegraphics[width=0.6\columnwidth]{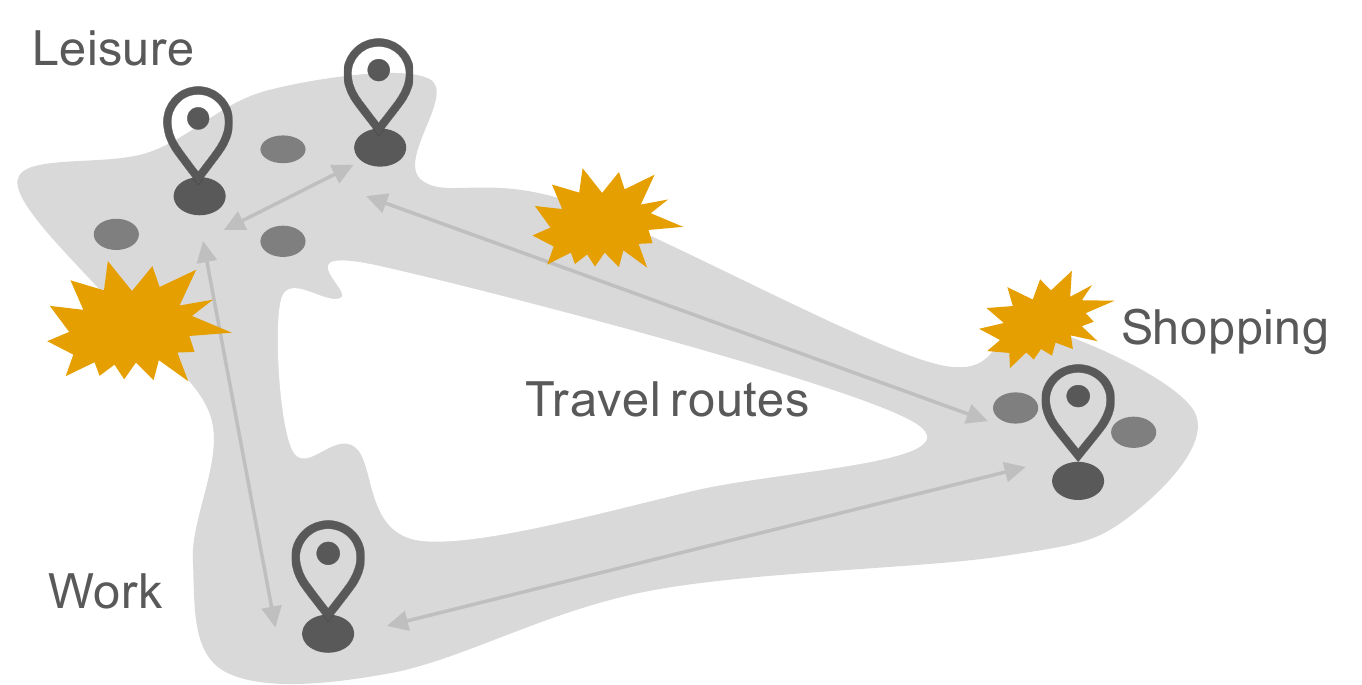}
\caption{Schematic visualization of crime pattern theory. According to this representation, crime (\ie, orange stars) emerges where the daily awareness spaces of victims and offenders overlap (\ie, gray shading).}
\label{fig:awareness_space}
\end{figure}

One of the few theories that specifically accounts for spatio-temporal variations due to human mobility is \emph{crime pattern theory} \cite{Brantingham1993}. According to this theory, people pursue a variety of recurrent activities -- such as working or shopping -- at different locations, which are called \emph{routine activity nodes}. Around these locations, as well as on the \emph{pathways} between these locations, the theory defines so-called \emph{awareness spaces} (equally for offenders and for victims). Crime can potentially occur where and when these awareness spaces intersect. Hence, more crime should crop up in areas that either attract more people due to a larger number of routine activity nodes, or that are located along frequently traveled routes between these nodes. The idea is illustrated in \Cref{fig:awareness_space}, which presents routine activity nodes (\ie, work, shopping, and leisure) in an area, as well as the corresponding pathways. Accordingly, crime should theoretically occur in the represented awareness spaces. However, this theory has hitherto not been rigorously validated.

Testing crime pattern theory is challenging as it requires exact mobility patterns for determining awareness spaces in the form of both routine activity nodes and pathways. Another challenge is that testing crime pattern theory not only requires detailed mobility data, but, in order to obtain valid estimates, such data should be of high temporal granularity, \ie, at the level of hours. However, these are limitations of prior research, which has used only aggregated proxies for activity locations and has largely neglected the routes to/from these locations. For instance, \mycite{Kadar2018} use aggregated check-in counts from location technology platforms to account for ambient populations. However, we are not aware of any literature that links the mobility trajectories of individuals to spatio-temporal crime concentrations. Hence, it has remained unclear how to operationalize daily awareness spaces as represented by both routine activity nodes \emph{and} pathways.  


In this work, we test crime theory with reference to human mobility: we propose a novel approach for operationalizing awareness spaces consisting of both routine activity nodes and pathways in order to study the relationship between urban population flows and spatio-temporal crime concentrations. For this purpose, we suggest the use of data from large-scale online location services, \eg, Foursquare as in our work. Foursquare is a technology platform that powers leading business solutions and consumer products through a deep understanding of location. The anonymized data used in this study originates from the company's consumer apps, which allow users to report online the venues where they go for eating, drinking, working, and so forth (so-called "check-ins"). This provides us new insights into the spatio-temporal dynamics of human activity, which is beneficial for our research: venues associated with check-ins denote activity nodes, while transitions between the venues represent pathways. We later derive how we infer the pathways as the shortest path across the city between consecutive check-ins. According to crime pattern theory, more check-ins or pathways in a neighborhood should increase the number of overlapping awareness spaces and thus elevate crime levels. 


\textbf{Explanatory findings.} We find confirmatory evidence for crime pattern theory. That is, crime concentration in a neighborhood is described by check-in counts in the neighborhood venues \emph{and} by transitions where people only traverse the neighborhood without stopping. Both variables are positive and statistically significant, even when controlling for location, time, and past crime. These findings are repeated in three large urban areas with varying socio-demographic characteristics: San Francisco, Philadelphia, and Chicago. For instance, in San Francisco, 100 additional check-ins increase crime by \SI{4.77}{\percent} while, more importantly, 100 additional pass-through visitors en route to other locations result in an additional \SI{7.22}{\percent} increase. This highlights the relative importance of human pathways in explaining crime.  


\textbf{Predictive power.} In an out-of-sample setting, we tested the extent to which data from awareness spaces can improve crime forecasting in large cities. Here machine learning models incorporating human mobility flows improve predictive ability: for instance, the corresponding mean squared error~(MSE) over the baseline model of historical crime is reduced by \SI{30.98}{\percent} in the case of random forests applied to the San Francisco data. This improvement is larger than that obtained by accounting only for routine activity nodes, which corresponds to a \SI{28.86}{\percent} lower error over the baseline model in the same setup. Based on an analysis of the mean absolute error~(MAE), the improvement over the baseline model amounts to about 2,910 additional yearly crimes that could have been correctly predicted. Altogether, these findings suggest considerable benefits of human mobility data for the task of crime forecasting in public decision-making. 


\textbf{Contributions.} We advance computational social science in the following directions:
\begin{enumerate}
\item We investigate crime theory at scale. In fact, our findings provide strong evidence supporting crime pattern theory and suggest mixed evidence concerning the concept of humans acting as natural guardians. Furthermore, the findings hold true even when accounting for alternative explanations from social disorganization theory.
\item We establish the value of data from location technology platforms in understanding and predicting crime in large cities. For this, we leverage Foursquare data (in anonymized form) consisting of almost 14.8~million transitions and, as such, go beyond the pure check-in counts of earlier research \egcite{Kadar2018}. 
\item We discern between different mobility flows: we infer the number of people (i)~that are incoming or outgoing from a neighborhood, (ii)~transitioning between different venues within the neighborhood, or (iii)~that potentially pass through the neighborhood on their way to other areas. Altogether, this enables us to operationalize awareness spaces from crime pattern theory, namely both routine activity nodes and pathways between these nodes.  
\item We confirm the results in an extensive series of robustness checks (different cities, alternative computations of pass-through flow, alternative model specifications, and alternative predictors). In particular, the finding remains consistent across different urban regions, with different population characteristics. 

To this end, we employ existing computational approaches from network science and machine learning as we detail later in the methods section.

\end{enumerate}
\section{Related Work}

\subsection{Theoretical background on crime concentrations}


Studies in environmental criminology have described crime concentrations based on (a)~temporal and (b)~spatial dimensions as follows. Temporal variations in crime have been found across seasons, months, and even days of the week \cite{Andresen2013,Andresen2015}. Spatial variations of crime concentrations have been repeatedly observed at various spatial scales, such as areal units or at street level \cite{Johnson2010}. 


Factors that explain spatial variations in crime concentrations were initially located in traditional attributes of the resident population. For instance, it was found that concepts from social disorganization theory, such as residential stability, ethnic heterogeneity, and income levels, help in explaining neighborhood crime \cite{Sampson1997,Shaw1942}. An alternative explanation builds upon static land use indicators as metrics for neighborhood function \cite{Brantingham1995}. 
Later, crime pattern theory \cite{Brantingham1993} hypothesized that one source of spatio-temporal variations in crime concentrations could be due to human mobility, specifically crime occurs when the activity space of a victim or target intersects with the activity space of an offender; yet, this has not been rigorously tested.


There are different theories that offer opposing suggestions concerning how human presence in an area should relate to crime. On the one hand, crime pattern theory stipulates a positive relationship \cite{Brantingham1993}: greater population flows should increase the number of awareness spaces and thus result in elevated levels of crime. On the other hand, it has been argued that people on the street should act as guardians and thus lower crime \cite{Jacobs1961}. According to this eye-on-the-street concept, policy-makers can specifically elicit such behavior through higher densities and diversity of human activities.

\subsection{Geo-tagged data for modeling crime}


Many studies in environmental criminology use static characteristics of the spatial environment such as the presence of bars or highway exits \cite{Brantingham1995}, while most recent work builds on destination data from transportation surveys \cite{Felson2015}. The former approach has a temporal dimension by implication only, while the latter is prone to omitting tourist visitors and further lacks temporal dynamics.

Works in computational social science and data mining have modeled aggregated crime counts based on features derived from human-generated digital data with geo-tagging. Examples include points-of-interest \cite{Kadar2019,Wang2016}, Twitter data \cite{Gerber2014,Vomfell2018}, telecommunication data \cite{Bogomolov2014,Traunmueller2014}, transportation data \cite{Kadar2018,Wang2016}, and check-in counts from location services data \cite{Kadar2017,Kadar2018}. However, the latter work employed the number of check-ins as an indicator of the ambient population and is thus limited to pure check-ins, while ignoring actual movements. \mycite{Wang2016} experimented with aggregated taxi data with the goal of modeling connectivity between communities in the context of crime. However, it overlooks the actual trajectory of the journey such as being manifested in pass-through traffic. 


The modeling approaches in the aforementioned studies can be grouped as follows. First, the \emph{unit of analysis} (\ie, the spatial resolution of the crime variable) usually originates from discretizing the city, either into squared grid cells of a predetermined size \egcite{Gerber2014} or based on existing administrative units such as census tracts \egcitetwo{Kadar2018}{Vomfell2018}. We adhere to prior research by utilizing census tracts as our unit of analysis. Second, previous studies set the \emph{time resolution} according to crime counts that were aggregated on an annual \egcite{Wang2016}, monthly \egcite{Bogomolov2014}, or weekly basis \egcite{Vomfell2018}. In contrast, we follow a different path by choosing a more granular representation as this should better encode the fast pace of urban mobility; therefore, we advance by studying spatio-temporal patterns of crime and human flows based on an hourly resolution. Third, the \emph{model} varies depending on the objective of the study. If an empirical quantification of the effect strength was desired, then researchers utilized correlation analysis and simple association-based regressions as this facilitates theoretical hypothesis testing in order to better understand the historic relationship between data and crime \egcite{Traunmueller2014}. If decision support for public stakeholders was needed, then machine learning models were estimated in order to forecast future crime hotspots \egcitetwo{Bogomolov2014}{Kadar2018}. We adopt both strategies: we perform statistical tests in order to quantify the effect of mobility flows on crime; afterwards, we also test to what extent this relationship generalizes to out-of-sample evaluations concerning future crime in order to support public decision-makers in crime forecasting. 

\subsection{Mining data from online location services}


Data from online location platforms captures human activity at high resolution. Accordingly, this type of data was instrumental in basic research on human dynamics, where it helped in identifying universal patterns of human urban mobility \cite{Noulas2012} and in discovering dynamic clusters of urban activity \cite{Cranshaw2012}.


Given the aforementioned strengths of location service data, it is not surprising that research has leveraged it in a wide range of applications. For instance, it was used for measuring urban social diversity \cite{Hristova2016} and the impact of cultural investments \cite{Zhou2017}. Other works apply this type of data in a business context, such as retail site location planning \cite{Karamshuk2013} or for predicting the survival of retail stores \cite{DSilva2018}. Even other works adapt it to a healthcare setting, \eg, when modeling the spatio-temporal evolution of chronic diseases \cite{Wang2018}. However, mobility derived from check-ins data has not yet been tested in relation to crime. 

\subsection{Research gap}

Our research differs from existing works with respect to the following dimensions, which are addressed for the first time in our study: (i)~we test crime pattern theory at scale; (ii)~we model crime based on human flows inferred from online location services; and (iii)~we discern the relative importance of routine activity nodes and pathways as separate elements of awareness spaces. 

\begin{figure*}[htbp]
\centering
\footnotesize
\begin{tabular}{ccc}
(a)~Crime counts & (b)~Check-in counts & (c)~Pass-through flow \\
\includegraphics[width=0.3\textwidth]{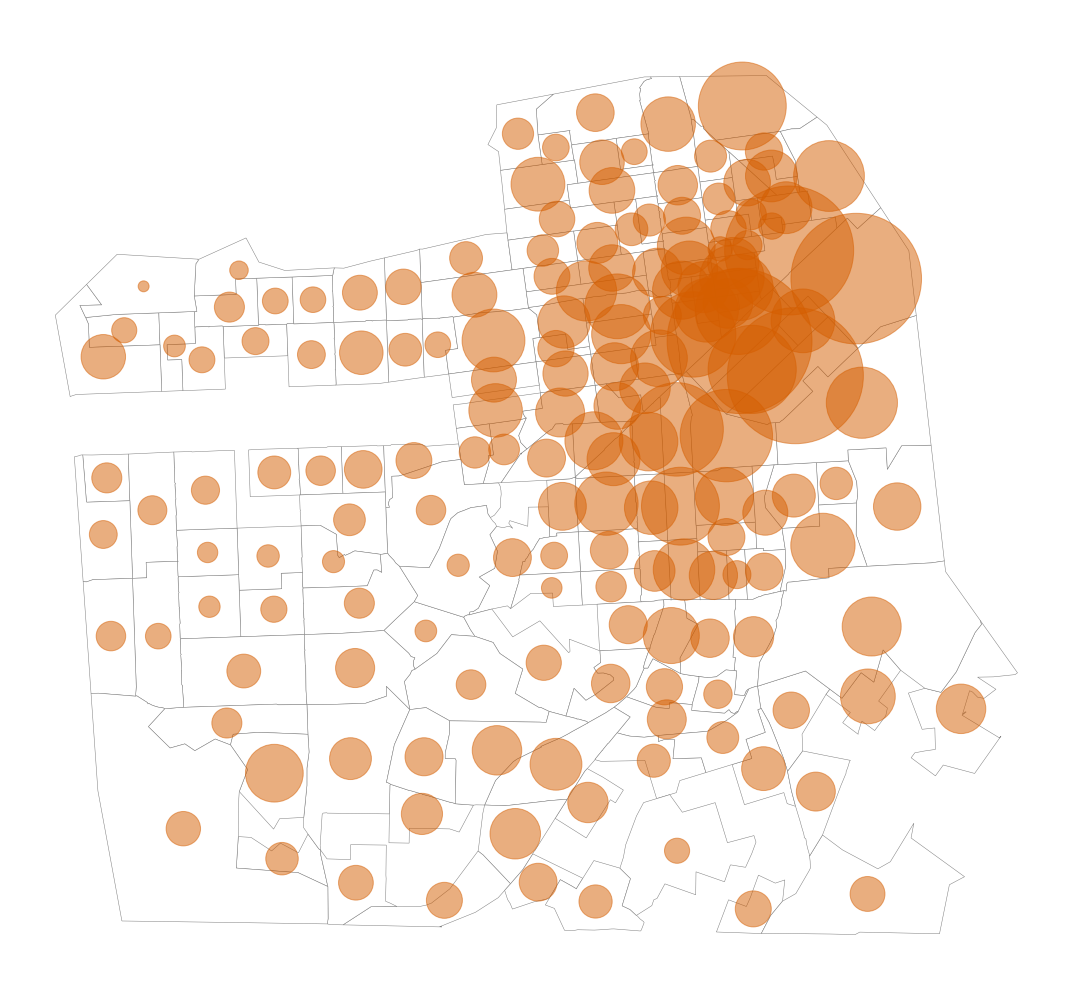} &
\includegraphics[width=0.3\textwidth]{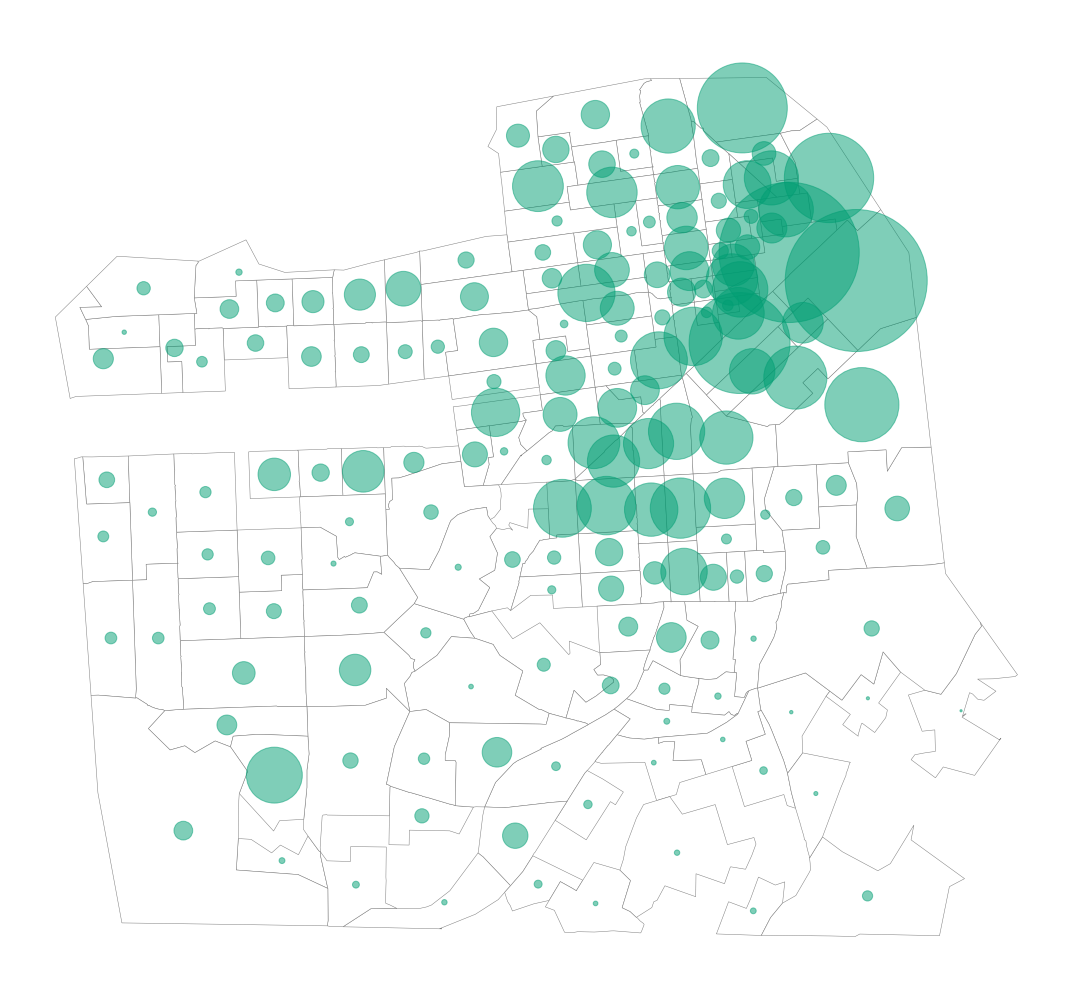} &
\includegraphics[width=0.3\textwidth]{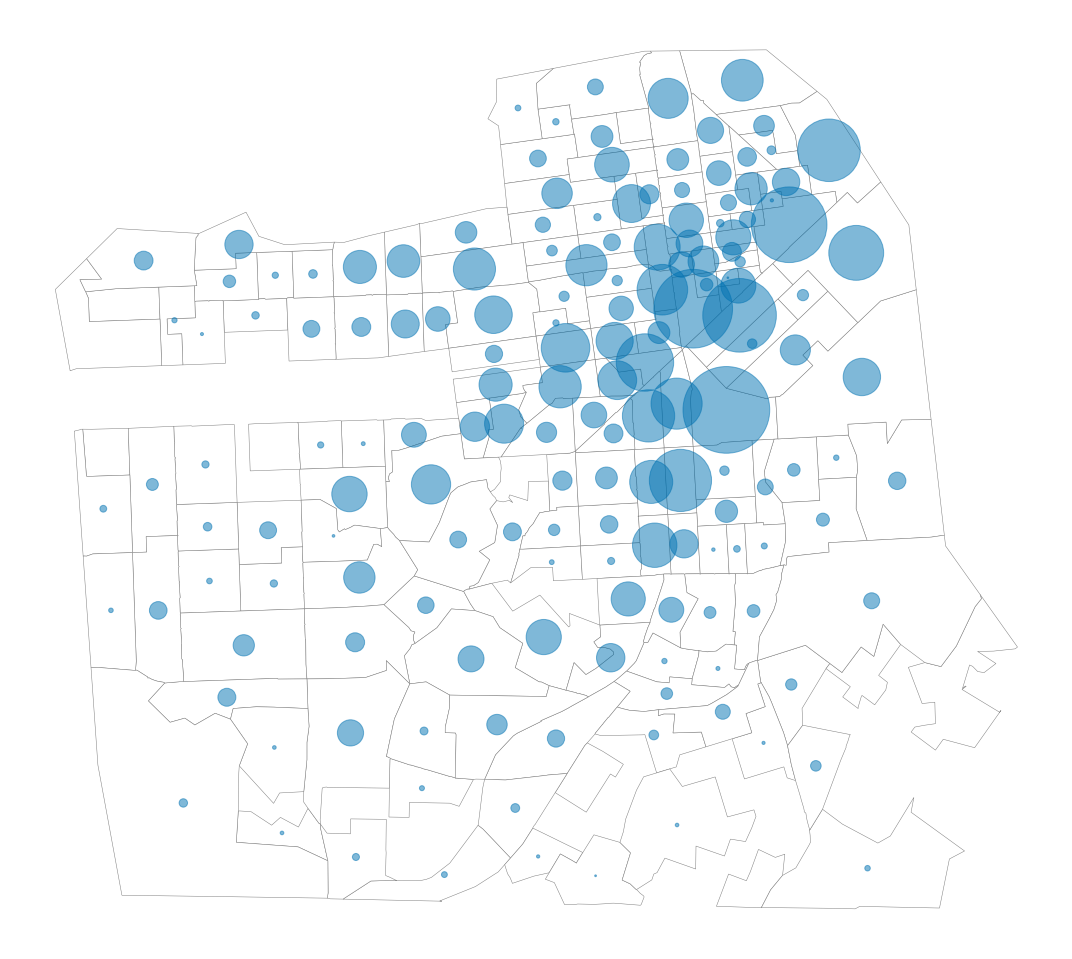} 
\end{tabular}
\caption{Descriptive statistics comparing San Francisco neighborhoods (using 2012 data). Crime counts are reported in original units, while check-ins and pass-through transitions are in 100s. These values are indicated by the size of the circles; the background refers to the corresponding census tracts.}
\label{fig:city_map}
\end{figure*}

\section{Data}


Our research is based on the following empirical setting. We study crime concentrations in detail over the course of two years in the city of San Francisco. Further, we repeat the main experiments on data from Philadelphia and Chicago. As such, we cover several U.S. urban areas of varying sizes, and with different socio-demographic characteristics and levels of technical savviness. Our choice of locations is motivated by the fact that crime is more concentrated and a more pressing problem in dense urban environments \cite{Bettencourt2007}.


By studying hourly temporal profiles of crime and mobility, the data is eventually aggregated to a study horizon $T = \{ 0, \ldots, 167 \}$ that comprises each hour of the week. 


We study neighborhood crime at the level of individual census tracts.\footnote{Throughout this paper, we use the two terms interchangeably. Census tracts are administrative units defined by the United States Census Bureau and are home to 4,000 inhabitants on average.} The total number of census tracts is 196, 384, and 397 respectively, of which we discard those with a population of fewer than 100 inhabitants and those below a certain threshold of annual check-ins (\ie, 100). This results in a final set $V$ of 169, 225, and 280 census tracts, respectively.

\subsection{Crime dataset} 

The crime data originates from police reports that refer to crime incidents committed in same time period as our Foursquare dataset. The data is publicly available and can be downloaded from the Open Data platforms of the three cities.\footnote{\tiny{\url{data.sfgov.org}, \url{www.opendataphilly.org}, \url{data.cityofchicago.org}}}. As was done in prior literature \cite{Kadar2018}, we consider crime incidents involving one of five types of felonies, namely larceny/theft, robbery, assault, burglary, and vehicle theft. All crime incidents were aggregated at census tract level. In total, this resulted in a final sample of almost 115,000, 148,000, and 158,000 crimes, respectively.

\Cref{fig:city_map}(a) compares the distribution of crime across San Francisco's census tracts, showing a particularly high concentration in downtown.

\subsection{Foursquare dataset with user transitions} 


Our data from Foursquare contains the complete\footnote{For privacy reasons, transitions from/to home locations were removed by Foursquare before export. In our case, it allows us to focus on all routine activities that are non-residential.} and anonymized\footnote{Personal identifiers were removed by Foursquare before export.} sample of user transitions recorded from January 2012 to December 2013 in the three cities. Overall, the dataset counts almost 2.7 million, 2.2 million, and 5.5 million transitions in each city, respectively. The dataset was granted through a collaboration between Foursquare and the University of Cambridge. IRB approval was obtained in order to work on the data for the purpose of this study. 


Our dataset has several advantages. Unlike raw check-in data, our dataset provides transitions that track the movements of individual users between activity nodes and therefore enables us to eventually infer mobility flows. Also, because we used a database export, our sample is not subject to the usual biases that arise during data collection when using APIs or scraping techniques. Nevertheless, we should point out that, by making no specific assumptions concerning the nature of the data, the findings should generalize to other sources of location data, such as explicit or implicit check-ins from platforms such as Google or WeChat.

On the downside, the general limitations of data captured in online services also apply to Foursquare, and we acknowledge them in the discussion section. 


An anonymous transition is defined as a pair of consecutive check-ins recorded by the same user at two different venues within a timespan of three hours. A transition is essentially a tuple with additional information, namely start time, end time, source venue, and destination venue. 


Each Foursquare venue belongs to a category specified according to a detailed taxonomy. We use this information to assign each check-in to one of the following main routine activity types: work/study (\eg, \emph{offices}), restaurants/bars (\eg, \emph{Spanish restaurants}), leisure (\eg, \emph{outdoors and recreation} and \emph{entertainment}), shopping (\eg, \emph{shops}), and travel (\eg, \emph{airports}). With the exception of the restaurants/bars activity, the check-ins are evenly distributed among the five activity types: work/study (\SI{10.11}{\percent}),restaurants/bars (\SI{50.11}{\percent}), leisure (\SI{12.82}{\percent}), shopping (\SI{15.56}{\percent}), and travel (\SI{11.40}{\percent}).

Finally, each venue comes with longitude and latitude coordinates that we can use to identify the corresponding census tract. \Cref{fig:city_map}(b) illustrates the volume of check-ins across different census tracts, while \Cref{fig:city_map}(c) reports the pass-through visitors derived from the transitions as detailed in the next section.

\begin{figure*}[htbp]
\centering
\footnotesize
\begin{tabular}{ccc}
(a)~Spatial adjacency network & (b)~Origin-destination network & (c)~Shortest-path network \\
\includegraphics[width=0.3\textwidth]{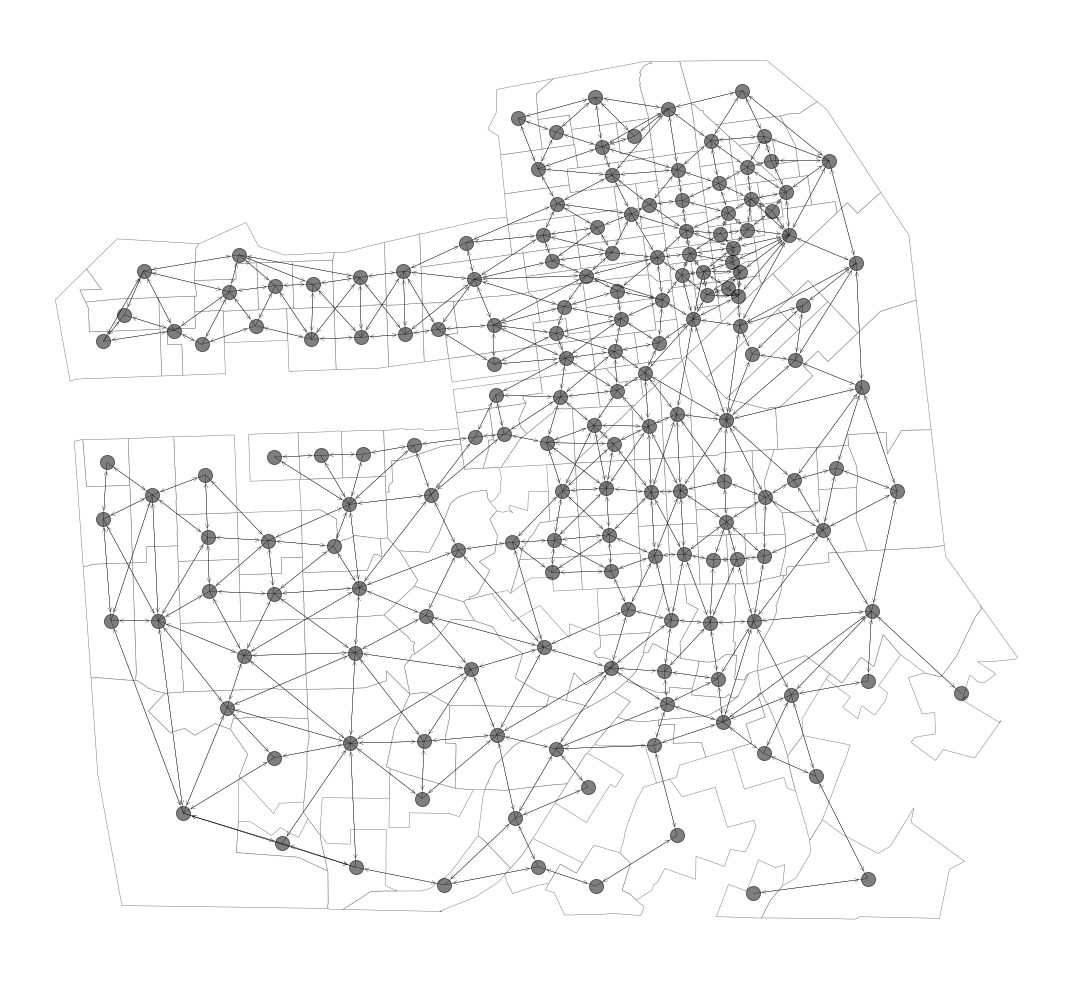} &
\includegraphics[width=0.3\textwidth]{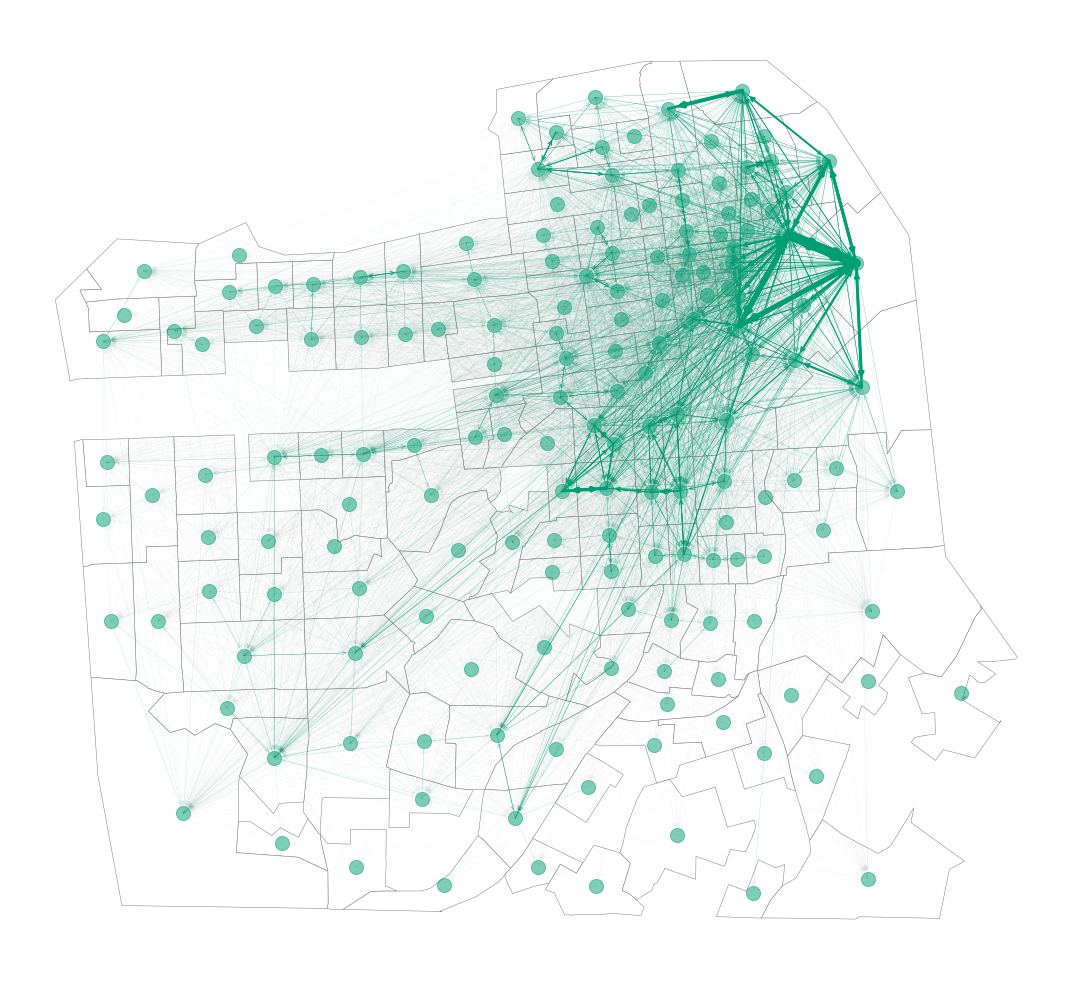} &
\includegraphics[width=0.3\textwidth]{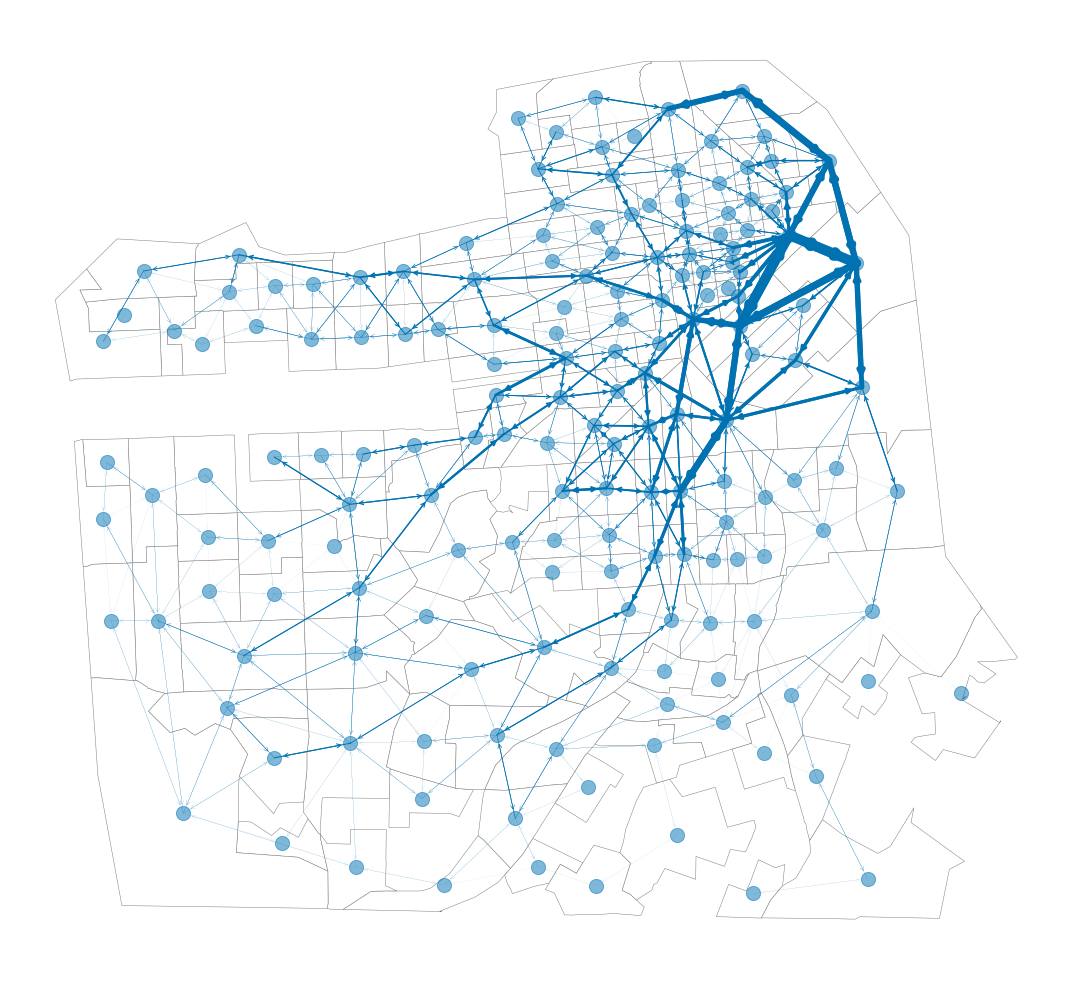} \\

\includegraphics[width=0.3\textwidth]{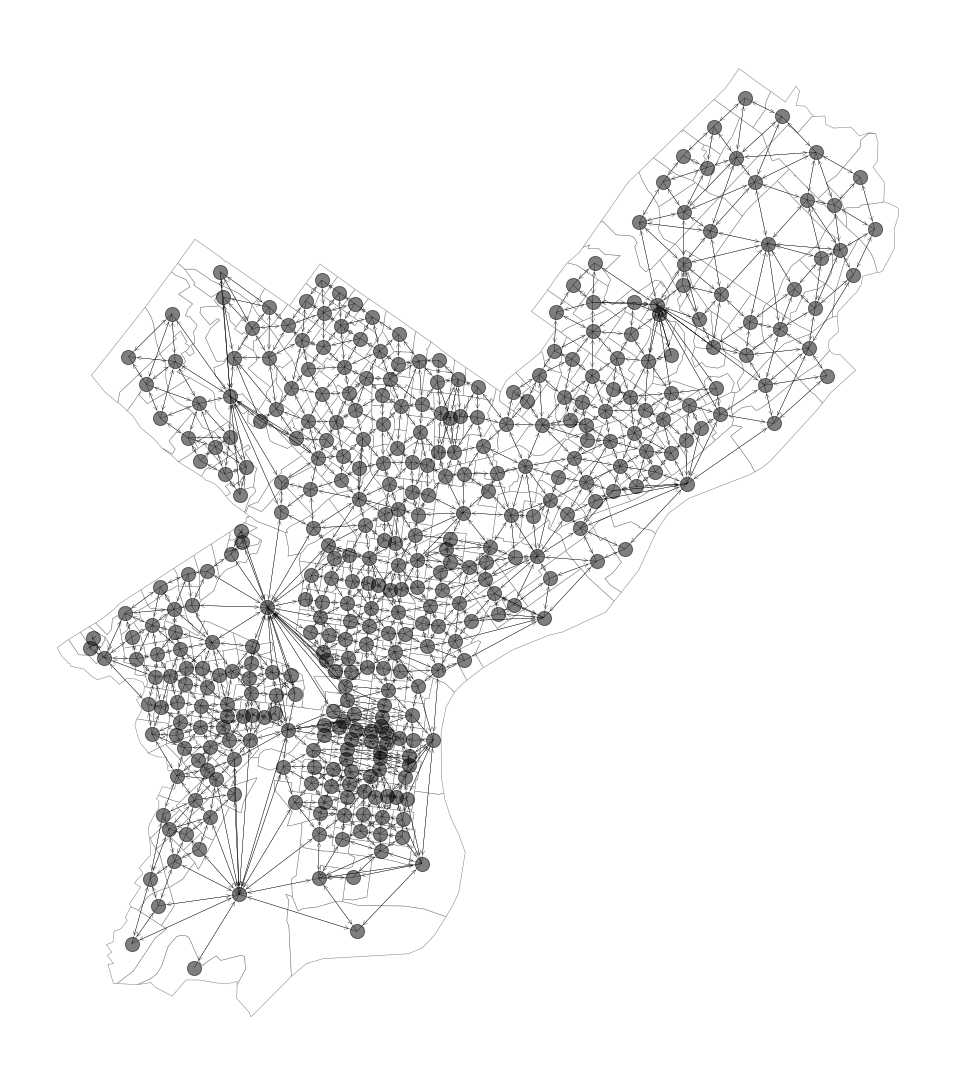} &
\includegraphics[width=0.3\textwidth]{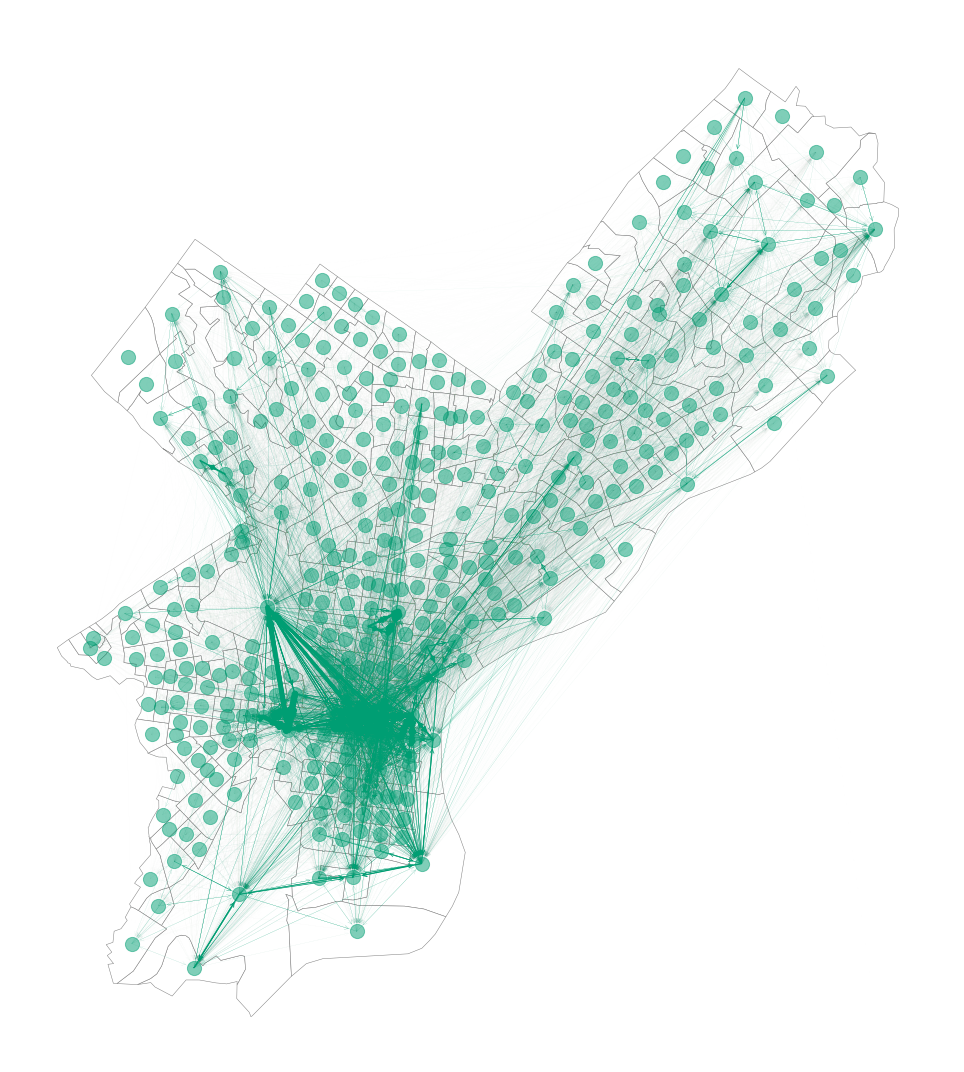} &
\includegraphics[width=0.3\textwidth]{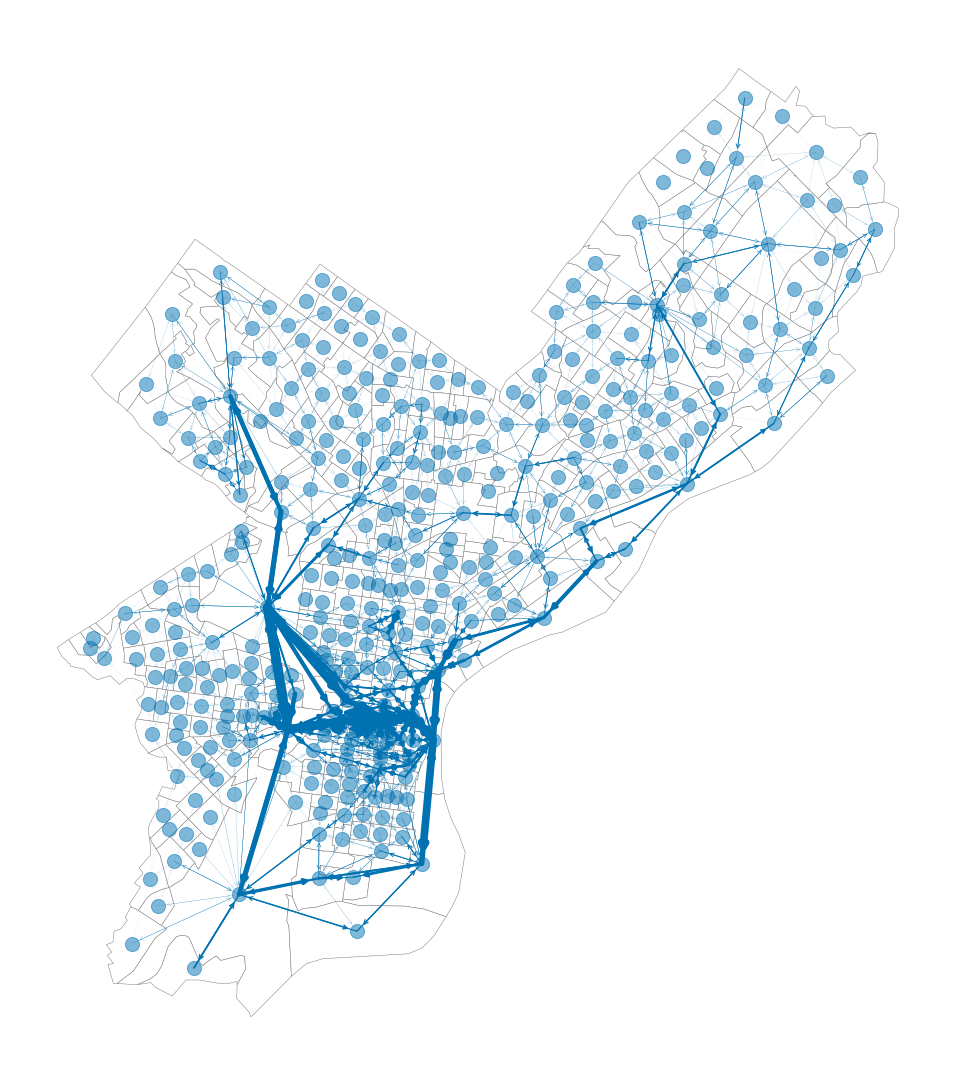} \\

\includegraphics[width=0.3\textwidth]{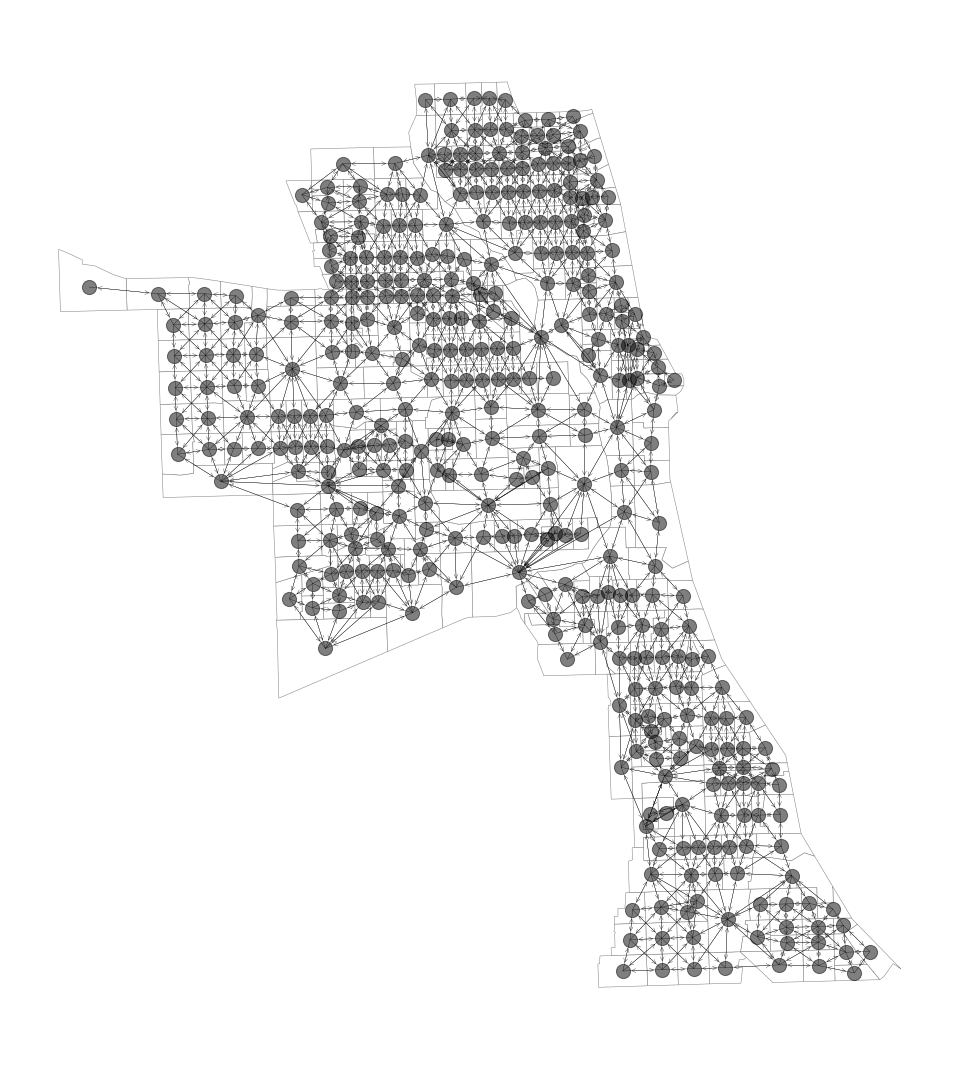} &
\includegraphics[width=0.3\textwidth]{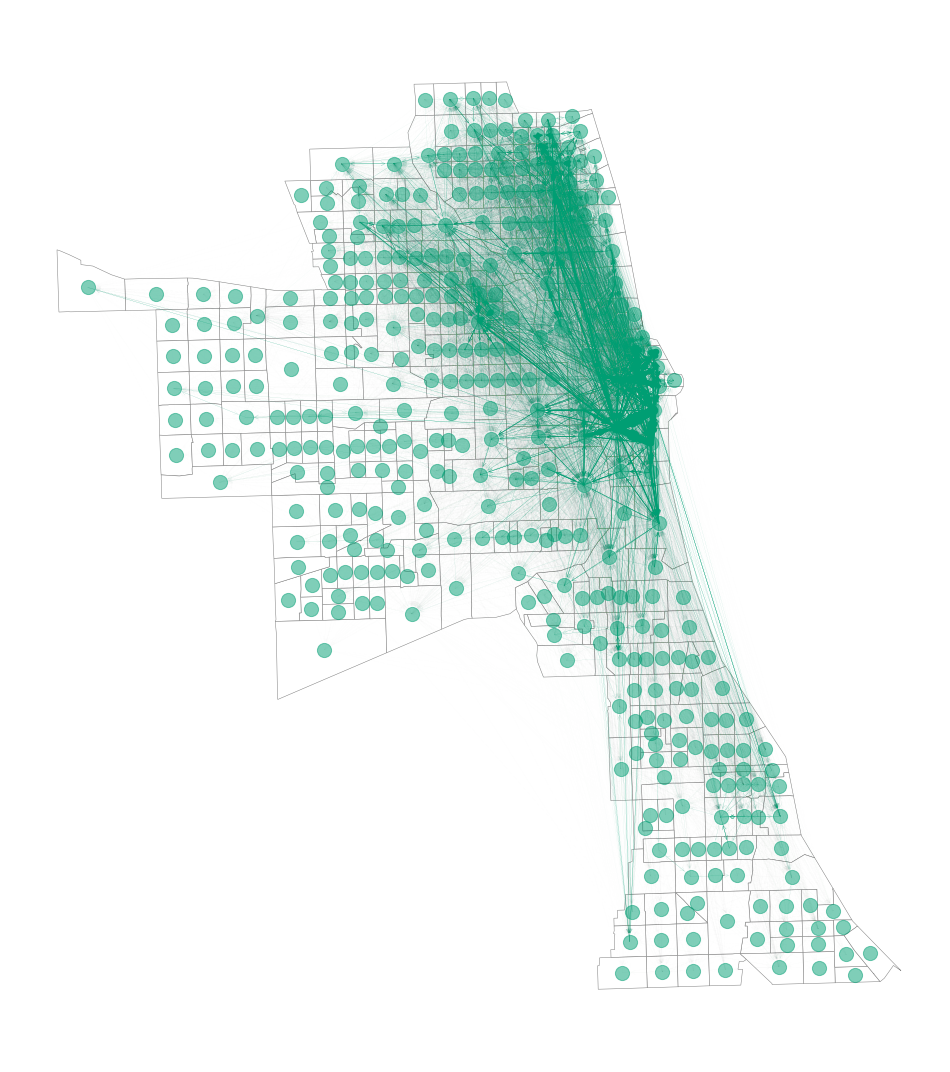} &
\includegraphics[width=0.3\textwidth]{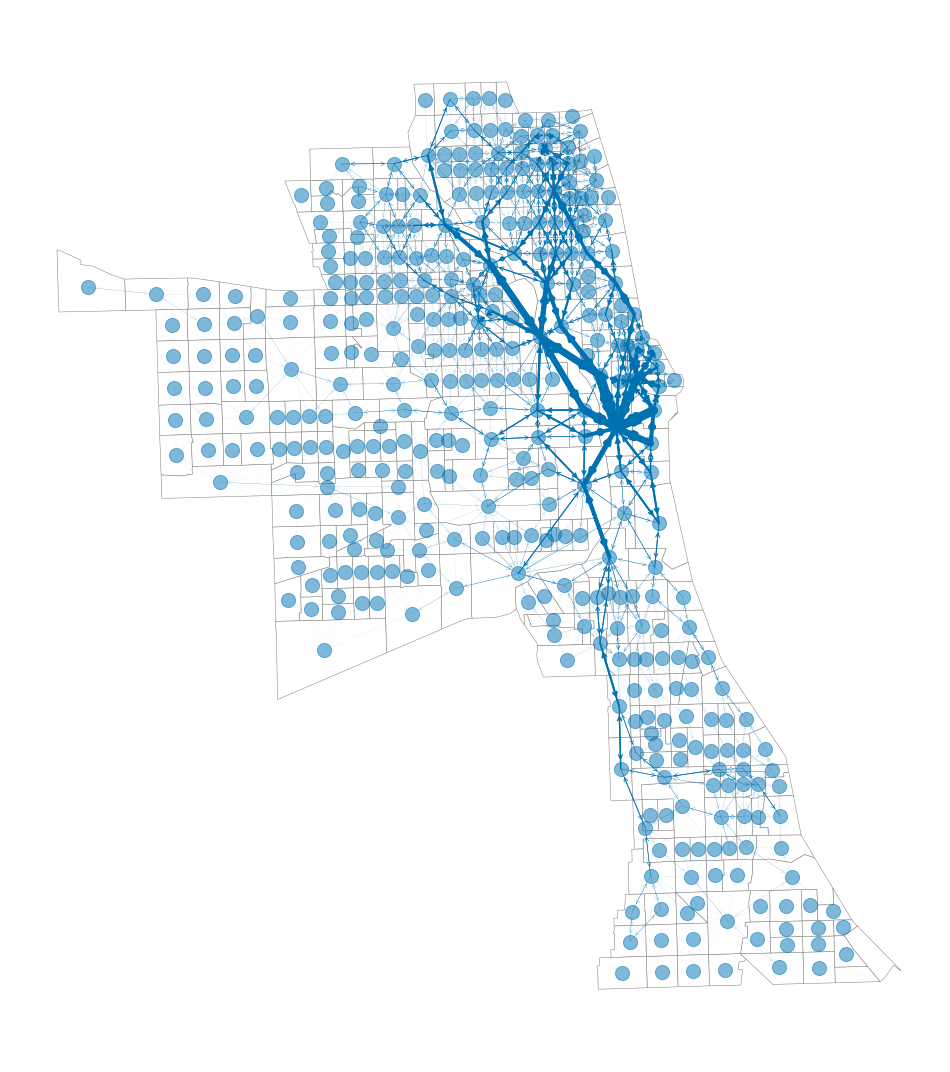} \\
\end{tabular}
\caption{Visualization of computational steps in order to retrieve the pass-through flows in San Francisco, Philadelphia, and Chicago (based on 2012 data). The line weight corresponds to the number of transitions; the background refers to the corresponding census tracts.}
\label{fig:network}
\end{figure*}

\section{Methods}

\subsection{Feature generation} 

Our objective is to operationalize awareness spaces by decomposing transitions at census tract level into the following variables:
\begin{itemize}
\item \emph{Check-ins.} These refer to the routine activity nodes and, in our case, we simply count the number of total check-ins, \ie, $\mathit{checkins}_{i,t}$, per neighborhood $i \in V$ and hour $t \in T$ in our dataset. We later split this variable further into \emph{incoming/outgoing transitions} and \emph{self-loop transitions}, whereby a person remains in the same neighborhood. In the former incoming/outgoing transitions, we do not distinguish whether an individual enters or leaves a neighborhood; instead, we are only interested how often an awareness space appears due to node activity. Given that we later introduce a graph-based notation, we refer to the latter variable, the transitions within the same neighborhood, as self-loop transitions since both origin and destination correspond to the same neighborhood node.
\item \emph{Pass-through flows.} This variable represents all transitions that potentially cross a neighborhood, but where origin and destination are located elsewhere. Hence, we need to infer the likeliest path that a person undertook when moving from origin to destination. 
\end{itemize}
Note that all variables are calculated at hourly level for each census tract. 

Computing pass-through flows requires additional processing steps that we detail in the following section. It is based on the idea of modeling the city as a network in which nodes represent neighborhoods. Then we replace a transition from the location service data with the shortest path in the neighborhood network. 

\subsection{Computation of pass-through transitions} 

We now introduce a network-based formalization of transitions that later aids the computation of pass-through flows. In the following, we make the simplified assumption that people can move between two neighborhoods if they simply share a common border. Later in the paper, we test another specification where traveling in a city must follow existing transportation lines, yielding consistent findings. We translate neighborhoods into nodes $V$ and further let edges $E$ connect adjacent neighborhoods. These then form a \emph{spatial adjacency network} $G_{\text{adj}} = (V,E_{\text{adj}})$. Formally, the location of a neighborhood $i \in V$ is modeled by the centroid location of the corresponding census tract. Two nodes $i,j \in V$ are connected by an unweighted edge $e_{ij}  \in E_{S}$ if and only if the perimeters of the corresponding areal units of the census tracts touch at one point (or more). The actual computation is based on queen's spatial contiguity.\footnote{Implemented in the \texttt{pysal} Python package.} The network is undirected and, hence, $e_{ij} \in V \Leftrightarrow e_{ji} \in V$. 


We compute pass-through flows by modeling the route that individuals have undertaken based on the network structure $G_\text{adj}$. For this purpose, let $\mathit{orig}$ and $\mathit{dest}$ refer to the origin and destination neighborhoods of a transition. Our goal is then to come up with a sequence of neighborhoods $n_{orig}, n_1, \ldots, n_{dest} \in V$ that defines the order of transitions between adjacent neighborhoods in order to reach $\mathit{dest}$ when starting in $\mathit{orig}$. Based on the spatial adjacency network, we first introduce an origin-destination network that encodes all potential transitions and, subsequently, derive a shortest-path network that yields the pass-through transitions. 


\emph{Step 1:} The \emph{origin-destination network} $G_\text{OD} = (V, E_\text{OD}, W_\text{OD})$ comprises all transitions that occur between different neighborhoods. This network is now directed and weighted: an edge $e_{ij} \in E_\text{OD}$ is added in the network if at least one transition occurred from neighborhood $i \in V$ to $j \in V$, while its weight $w_{ij} \in W_\text{OD}$ denotes the number of such transitions from $i$ to $j$. 


\emph{Step 2:} The origin-destination network quantifies the overall mobility flows, but it does not provide the routes. Hence, we compute a \emph{shortest-path network} where transitions are replaced by the shortest path between origin and destination according to the spatial structure in the spatial adjacency network. Using shortest paths in modeling transitions is a well-established tool in network science and is widely employed, for instance, in the computation of centrality metrics such as betweenness centrality \cite{Freeman1977}. 


The shortest-path network is computed as follows. Let $G_\text{SP} = (V, E_\text{SP}, W_\text{SP})$ denote the shortest-path network with directed edges $E_\text{SP}$ and corresponding weights $W_\text{SP}$. For each edge in $G_\text{OD}$, the shortest path $\tau$ according to the network structure in $G_\text{adj}$ is computed.\footnote{Implemented in the \texttt{pathpy2} Python package.} Then, for each edge along the path $\tau$, an edge is added to the shortest-path network $G_\text{SP}$. Its weight is set so that it is the sum of the weights from all shortest paths passing through it. The pseudocode for the previous steps is listed in \Cref{alg:gsp}, which returns the directed weighted network $G_\text{SP}$. 
 
\begin{algorithm}
  \caption{Construction of shortest-path network}
\footnotesize
  \begin{algorithmic}
    \State Initialize neighborhood nodes $V$ and set $E_\text{SP} \gets E_\text{adj}$ 
    \For {each $e_{ij} \in E_\text{SP}$}:
    		\State Initialize edge weights to $w_{ij} \gets 0$
    \EndFor 
    \For {each weighted transition $t_{kl} \in E_\text{OD}$}: 
    	 \State Compute the directed shortest path $\pi_{kl}$ between nodes $k$ and $l$ in the network $G_\text{adj}$
	 \For {each directed edge $e_{ij} \in \pi_{kl}$}:
	  \State Increment the weight $w_{ij}$ by the number of transitions for $t_{kl}$ between $k$ and $l$; \ie, $w_{ij} \gets w_{ij} + \text{weight}[t_{kl}]$
	\EndFor 
    \EndFor \\
   \Return $G_\text{SP} = (V, E_\text{SP}, W_\text{SP})$ with $W_\text{SP} = \left\{ w_{ij} \right\}$
  \end{algorithmic}
  \label{alg:gsp}
\end{algorithm}

\begin{figure*}[htbp]
\centering
\footnotesize
\begin{tabular}{ccc}
(a) Census tract 12800: & (b) Census tract 11700: & (c) Census tract 10100:\\
Lombard Street & Union Square & Fisherman's Wharf \\
\includegraphics[width=0.3\textwidth]{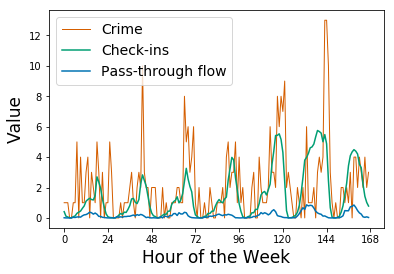} &
\includegraphics[width=0.3\textwidth]{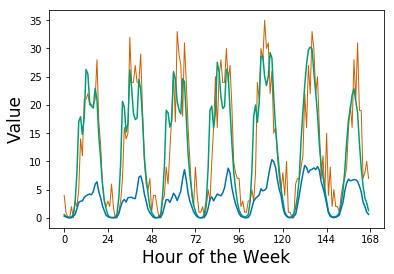} &
\includegraphics[width=0.3\textwidth]{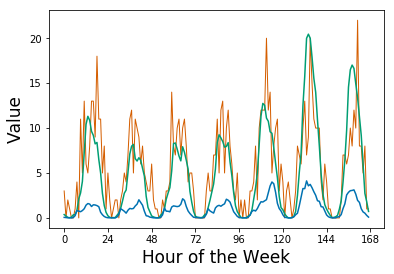}
\end{tabular}
\caption{Hourly temporal profiles reveal co-movements of the numbers for crime and mobility flows in San Francisco (shown: 2012 data). Crime counts are reported in original units, while check-ins and pass-through transitions are in 100s.}
\label{fig:temporal_profiles}
\end{figure*}

\subsection{Model specification} 


We perform a two-pronged evaluation whereby we vary the model depending on our objective; see \mycite{Breiman2001} for a detailed discussion. On the one hand, we perform an \emph{explanatory validation} with the aim of theory testing. Hence, we perform statistical tests in order to quantify how well regressors with human mobility data describe the dependent variable of interest, \ie, crime. As a result, one then interprets the estimated effect strengths, performs significance testing, and studies the explained variance based on historic (\ie, in-sample) data. On the other hand, \emph{predictive modeling} evaluates the ability of human mobility in forecasting future, unseen crime (\ie, out-of-sample). It is thus concerned with providing decision support to public stakeholders. By pursuing both explanatory and predictive modeling, we contribute findings of relevance to different audiences.

\subsubsection{Empirical modeling for theory testing.} 


For understanding how the regressors explain crime, we draw upon a panel generalized linear model~(PGLM) with location and time fixed-effects.\footnote{We used the implementation provided in the R~package \texttt{pglm}.} This model is suited to our research question for several reasons. First, the dataset consists of time series. Second, we must account for between-neighborhood and temporal heterogeneity by incorporating fixed-effects at both census tract level and for each of the 168 hours within a week. This is a common approach in urban computing \cite{Zhang2016}, since it specifically controls for unobservable spatio-temporal dynamics. Thereby, we control for the fact that some neighborhoods are more popular than others due to spatial characteristics, as well as the fact that crime across the whole city is subject to the same temporal variations. As the dependent variable, \ie, crime counts, exhibits an over-dispersed distribution with higher variance than mean, we follow earlier research \cite{Osgood2000,Wang2016} and model crime as a negative binomial distribution. 

The resulting PGLM is then formalized by 
\begin{align}
\label{eqn:PGLM}
\begin{split}
	&\mathit{crime}_{i,t} = \nu + \alpha_i + \theta_t + \beta \times \mathit{past\_crime}_{i,t}   \\ 
	              & + \gamma \times \mathit{check\textnormal{-}ins}_{i,t}+ \delta \times \mathit{pass\textnormal{-}through\_flow}_{i,t} 
\end{split}
\end{align}	
with intercept $\nu$ and coefficients $\beta$, $\gamma$, and $\delta$, where $i$ refers to the census tracts in $V$ and $t \in \{ 0, \ldots, 167 \}$ to all hours of the week. Eq.~\eqref{eqn:PGLM} contains fixed-effects for both neighborhood ($\alpha_i$) and time ($\theta_t$). Over-specification was avoided as one of the dummies for each dimension was automatically removed. We further incorporate historic crime counts from the previous year (\ie, 2011 when estimated with 2012 data) in the same neighborhood as an additional regressor, since historic crime profiles indicate whether a neighborhood was previously prone to crime. The coefficients $\gamma$ and $\delta$ then measure the extent to which both check-ins and pathways relate to crimes. 


We later experiment with several model variations.  We report the Akaike information criterion (AIC) as it should indicate which model should be preferred. First, we test several models where we omit the pass-through flow. Second, we extend Eq.~\eqref{eqn:PGLM} by splitting the total number of check-ins into: incoming/outgoing movements, on the one hand, and self-loops where people remain in the same neighborhood on the other hand. 

\subsubsection{Predictive modeling for crime forecasting.}

We further study the prognostic capacity of mobility flows in forecasting future crime. Here we employ random forest~(RF) and elastic net~(EN) models.\footnote{We used the implementation provided in the Python~package \texttt{scikit-learn}.} These are known in the literature for their ability to yield competitive prediction quality and mitigate overfitting by their built-in regularization.


In the predictive model, we choose a different approach for handling spatio-temporal heterogeneity as in the above PGLM. In fact, leveraging dummies would result in overfitting and compromise the prognostic capacity of the model. Still, crime is subject to considerable spatio-temporal heterogeneity/autocorrelation. Hence, we include spatial features: the longitude $x_i$ and latitude $y_i$ of neighborhood $i$. Formally, the location was derived from the centroid corresponding to the neighborhood. Temporal variability was modeled via ${hour}_t$ for each hour. Moreover, we create an additional binary feature to distinguish between weekdays and weekend. These features were employed together with all features involving crime and mobility flows. 


The estimation procedure is as follows. The models were built with variables analogous to the PGLM. All model specifications were trained based on data from 2012. The optimal hyper-parameters (number of trees, depth of trees, and number of features to consider in the node splits for random forest; the penalty constant and the mixing parameter between L1- and L2-regularization for elastic net) were selected automatically using 5-fold cross-validation. Afterwards, we compute the out-of-sample performance in evaluating hourly crime data throughout 2013. Across all experiments, we report the mean squared error~(MSE).

\section{Results}

\subsection{Descriptive statistics}


\Cref{fig:network}(a) depicts the location of neighborhoods and the corresponding spatial adjacency network. \Cref{fig:network}(b) shows all transitions between origin and destination. This is then translated into the shortest-path network from \Cref{fig:network}(c), which yields the pass-through flow. We do this for San Francisco as our prime analysis and, as a robustness check, for Philadelphia and Chicago.


\Cref{fig:temporal_profiles} compares the temporal profiles of crimes and mobility flows in San Francisco. The profiles are shaped by daily and weekly circadian variations of human activity. The graphs already reveal similarities between crime and mobility flows, which are investigated formally below.

\begin{table*}[htbp]
\centering
\footnotesize
\scalebox{0.8}[0.8]{
{
\begin{tabular}{ll S[group-separator={\,}] S[group-separator={\,}] S[group-separator={\,}]} 
\toprule
Model   &Regressors &{AIC (San Francisco)} &{AIC (Philadelphia)} &{AIC (Chicago)}\\
\midrule
Baseline PGLM &Past crime &85789.44 &103510.00 &121538.40\\
\midrule
PGLM (1a)  &Past crime, check-ins &85203.60 &103266.30 &121307.60\\
PGLM (1b) &Past crime, check-ins, pass-through flow & \bfseries 85078.51 & \bfseries 103081.20 &121287.00\\
PGLM (2a)  &Past crime, incoming/outgoing flow, self-loop flow &85141.41 &103283.40 &121113.00\\
PGLM (2b)  &Past crime, incoming/outgoing flow, self-loop flow, pass-through flow & 85094.53 &103114.30 & \bfseries 121092.20\\
\bottomrule

\end{tabular}
}
}
\caption{Model comparison for the explanatory approach. The model specifications follow the PGLM with both location and time fixed-effects from Eq.~\eqref{eqn:PGLM}. Crime from the previous year is included in all model specifications, but mobility features are varied. The best model in terms of AIC is highlighted in bold (estimates are based on 2012 data).} 
\label{table:PGLM_models}
\end{table*}

\begin{table*}[htbp]
\centering
\footnotesize
\scalebox{0.8}[0.8]{
{
\begin{tabular}{ll SSSSSS}
\toprule
Model   &Predictors &\multicolumn{2}{c}{MSE (San Francisco)} &\multicolumn{2}{c}{MSE (Philadelphia)} &\multicolumn{2}{c}{MSE (Chicago)}\\
\cmidrule(rrr){3-4}  
\cmidrule(rrr){5-6} 
\cmidrule(rrr){7-8}  
        &           & {Value} & {Improv. (\%)}           & {Value} & {Improv. (\%)} & {Value} & {Improv. (\%)}\\
\midrule
Historical   & {---} & 4.765 & {---} &2.536& {---}  &2.417 & {---}\\

\midrule
RF (1a)  &Past crime, check-ins                    &3.390	&28.86 &1.821		&28.19 &1.713 &29.13\\
RF (1b)  &Past crime, check-ins, pass-through flow &3.354	&29.61 &1.808		&28.71 &\bfseries1.450 &\bfseries40.01\\
RF (2a)  &Past crime, incoming/outgoing flow, self-loop flow                    &3.344	&29.82 &1.806		&28.79 &1.462 &39.51\\
RF (2b)  &Past crime, incoming/outgoing flow, self-loop flow, pass-through flow &\bfseries3.289	&\bfseries30.98 &\bfseries1.726	     &\bfseries31.94 &1.456 &39.76\\
\midrule
EN (1a)  &Past crime, check-ins                    &4.259	&10.62 &1.765	&30.40 &1.610 &33.38\\
EN (1b)  &Past crime, check-ins, pass-through flow &4.216	&11.52 &1.764	&30.44 &\bfseries1.599 &\bfseries33.84\\
EN (2a)  &Past crime, incoming/outgoing flow, self-loop flow                    &4.212	&11.60 &1.768	&30.28 &1.637 &32.27\\
EN (2b)  &Past crime, incoming/outgoing flow, self-loop flow, pass-through flow &\bfseries4.189	&\bfseries12.09 &\bfseries1.738	&\bfseries31.47 &1.621 &32.93\\
\bottomrule
\end{tabular}
}
}
\caption{Predictive performance of mobility features in forecasting future crime. The models include separate features for location and time. Improvements over the historical crime profile are reported in \%. The best performance in terms of MSE is highlighted in bold (models are trained on 2012 data and tested on 2013 data).}
\label{table:pred_results}
\end{table*}

\subsection{Empirical relationship between mobility flows and crime}


\Cref{table:PGLM_models} performs a model selection based on AIC in order to compare the use of different variables across the three cities (a smaller AIC value is preferred).
We further employ likelihood-ratio tests to compare the goodness-of-fit of two competing models based on the ratio of their likelihoods and run several tests: (1a) vs baseline, (1b) vs. (1a), and (2b) vs (2a). Here the variants that include pass-through flows are always preferred, i.e., they are better at a statistically significant level ($p < .001$). Therefore, pass-through flows should be considered in the final model. We note that the alternative model specification with incoming/outgoing flow, self-loop flow, and pass-through flow is only marginally inferior in two cities to the specification based on check-ins and pass-through flow. More importantly, the AIC worsens considerably when omitting mobility features. This confirms the importance of pathways in describing crime concentrations.

\begin{figure}[htbp]
\centering
\includegraphics[width=1\columnwidth]{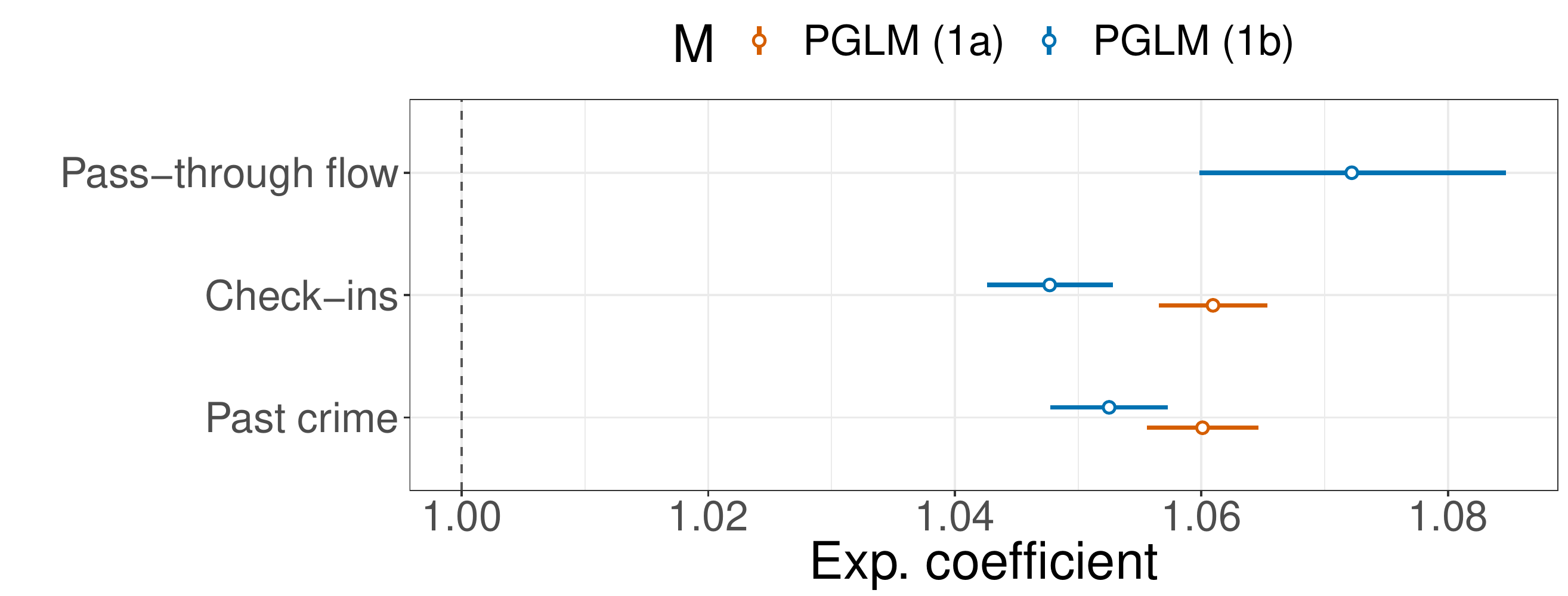}
\caption{How check-ins and pass-through flow describe crime concentrations based on San Francisco (2012 data). Visualized are exponentiated coefficients and the corresponding \SI{95}{\percent} confidence intervals from the PGLM model.}
\label{fig:PGLM_M23_coefs}
\end{figure}
\begin{figure}[htbp]
\centering
\includegraphics[width=1\columnwidth]{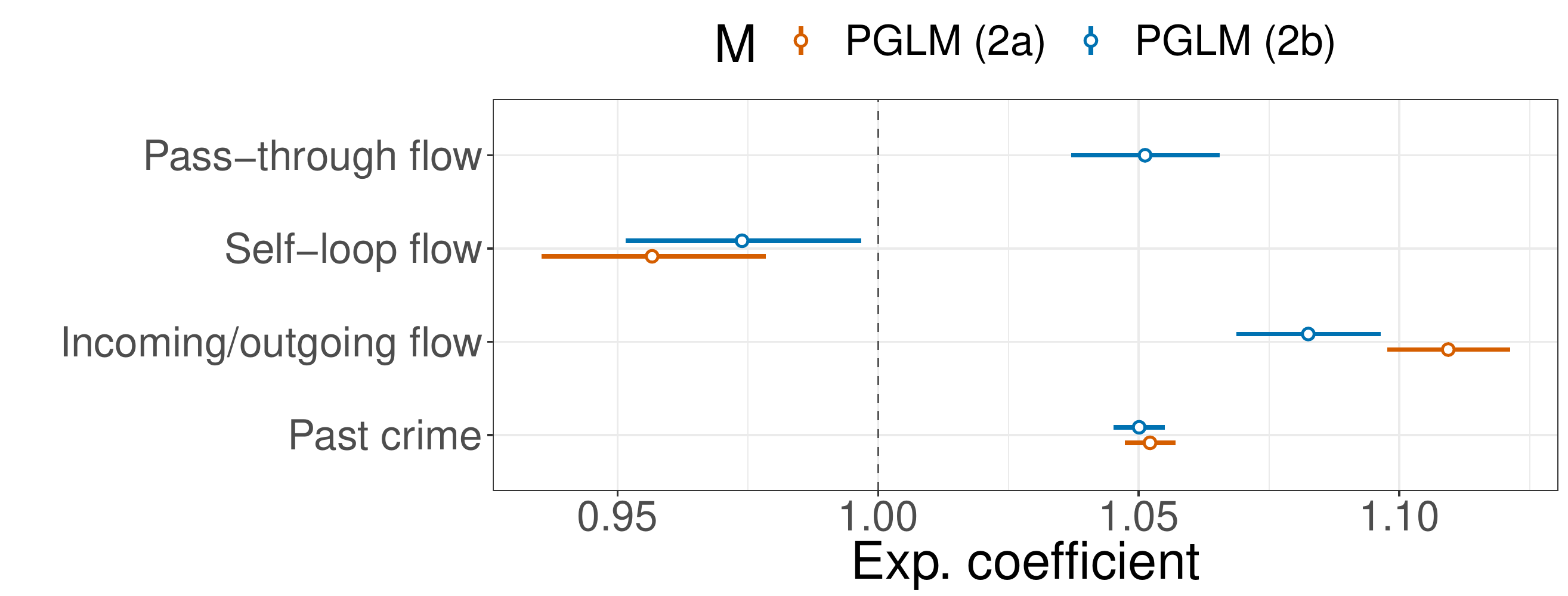}
\caption{How incoming/outgoing flow, self-loop flow, and pass-through flow describe crime concentrations based on San Francisco (2012 data). Visualized are exponentiated coefficients and the corresponding \SI{95}{\percent} confidence intervals from the PGLM model.}
\label{fig:PGLM_M45_coefs}
\end{figure}

Furthermore, all mobility coefficients are significant at the $\SI{0.1}{\percent}$ level, with the exception of the self-loop flow in PGLM~(2b) in San Francisco (significant at the $\SI{5}{\percent}$ level) and in Philadelphia (not significant). \Cref{fig:PGLM_M23_coefs,fig:PGLM_M45_coefs} report the detailed empirical results of the previous PGLMs in describing crime concentrations in San Francisco. The other areas under investigation yield similar results. We list different model specifications in order to confirm that the estimated coefficients remain robust. When interpreting the coefficients, we report so-called incident rate rations (\ie, the exponentiated value of the coefficient) as these measure the direct influence of a regressor on the outcome variable. 


We make the following observations from \Cref{fig:PGLM_M23_coefs}. As anticipated by crime pattern theory: both check-ins and pass-through flow are positively associated with crime. According to PGLM~(1b) in San Francisco, 100 check-ins suggest an increase in crime by \SI{4.77}{\percent} ($p < .001$). Yet we find an even larger effect for pass-through flow, where an additional 100 people transitioning through the neighborhood are linked to a \SI{7.22}{\percent} rise in crime ($p < .001$). Notably, the direction of these relationships remains robust across the other cities.


\Cref{fig:PGLM_M45_coefs} splits check-ins into both incoming/outgoing flow and self-loop flow. As the data shows, the awareness spaces introduced as a result of incoming/outgoing flow are linked positively with crime. However, self-loop flow, where people wander from one venue to another in the same neighborhood, is linked to reduced crime. According to PGLM~(2b) in San Francisco, every 100 such visitors correspond to a \SI{2.61}{\percent} decrease in crime counts ($p < .05$). This result is confirmed in all three cities. The implications for theory are discussed later.

\subsection{Predictive power of mobility flows in crime forecasting}

We now study the ability of mobility features to generalize to unseen observations and aid the prediction of future crime.  \Cref{table:pred_results} lists the out-of-sample prediction performance in all three cities.
We employ a benchmark in order to isolate the contribution of mobility features in forecasting: the historical profile which simply predicts $\mathit{past\_crime}_{i,t}$. In fact, related work suggested that crime clusters in space and time with regularity \cite{Johnson2010}, making historical crime a strong baseline. 

Overall, we notice higher improvements for the random forest models than for the elastic net models due to their non-linear nature.

In general, the best performance overall is achieved by the (2b) models using incoming/outgoing flow, self-loop flow, and pass-through flow. In the case of the random forest model in San Francisco, the model improves the MSE of the historical profile by \SI{30.98}{\percent}. It even outperform the equivalent model without pass-through flows, \ie, (2a), by 1.16 percentage points. We utilize a Wilcoxon signed-rank test to assess whether the results of the (2b) specifications are better than the results of the (2a) specifications and find that including path-through flow leads indeed to better results at a statistically significant level ($p < 0.001$)  across the different cities and models. This reveals the considerable prognostic value of pass-through flows in crime forecasting.

We also experiment with an alternative model specification that uses check-ins and pass-through flow; see (1b) models. According to the Wilcoxon test, they are always significantly superior to their corresponding (1a) specifications ($p < 0.001$). For instance, the inclusion of pass-through flows increases the performance by 0.75 percentage points when looking at the random forest results in San Francisco.

For conciseness, we report only the increase in prediction performance based on the MSE. We also compute alternative performance metrics, such as mean absolute error (MAE) and $R^2$, yet our findings still hold true. Based on the MAE performance improvements we can estimate how many more crimes would have been correctly predicted by a forecasting model including mobility features. In the case of San Francisco, RF~(2b) achieves a MAE $ = 1.181$, in comparison to a MAE $ = 1.386$ of the historical profile. Over 168 hours and 169 neighborhoods, this amounts to about 2,910 yearly crimes that could have been additionally correctly predicted.\footnote{Since, by just looking at MAE, we do not know whether the model under- or over-predicted the values, we took a conservative approach, and only considered half of the MAE improvement in the computation: $168 * 169 *(1.386 - 1.181) / 2 \approx 2,910$.}

\subsection{Sensitivity analysis by activity types}

\begin{figure}[htbp]
\centering
\includegraphics[width=1\columnwidth]{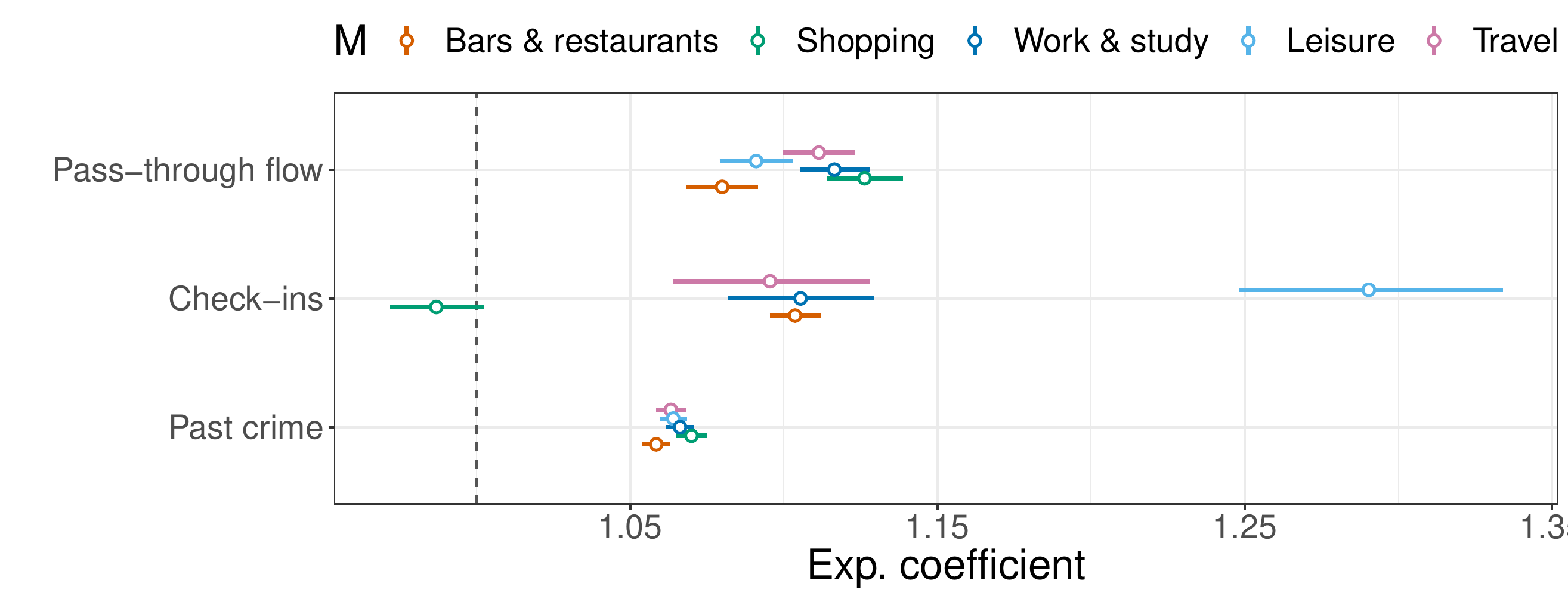}
\caption{How different types of check-ins describe crime concentrations based on San Francisco (2012 data). Visualized are exponentiated coefficients and the corresponding \SI{95}{\percent} confidence intervals from the PGLM model.}
\label{fig:PGLM_M3_activities_coefs}
\end{figure}

We further create specific instantiations of PGLM~(1b) in San Francisco where we only consider check-ins from a specific type of routine activities (\eg, only check-ins at bars and restaurants). As a result, we find that routine activities vary in their relationship with crime. Based on the results in \Cref{fig:PGLM_M3_activities_coefs}, we conclude that leisure activities have the highest association with crime, while shopping may be negatively linked to crime. Needless to say, the effect of pass-through flow remains positive and significant across all model specifications.

\subsection{Sensitivity analysis by crime types}

\begin{figure}[htbp]
\centering
\includegraphics[width=1\columnwidth]{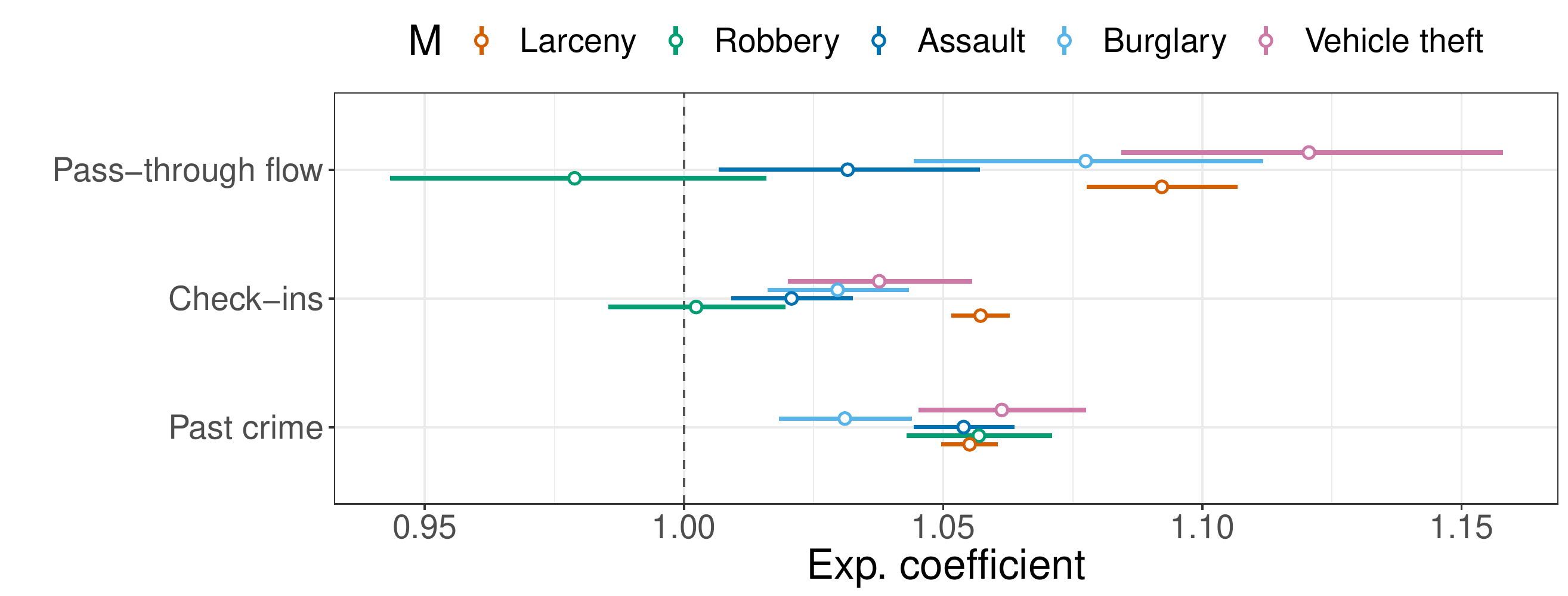}
\caption{How check-ins and pass-through flow describe different types of crime concentrations based on San Francisco (2012 data). Visualized are exponentiated coefficients and the corresponding \SI{95}{\percent} confidence intervals from the PGLM model.}
\label{fig:PGLM_M3_types_coefs}
\end{figure}

We also create specific instantiations of PGLM~(1b) in San Francisco where we consider as the dependent variable only incidents from a specific crime type: larceny, assault, burglary, robbery, and vehicle theft. The different types of crime have different relationships with the number of people spending time or passing through the neighborhood. \Cref{fig:PGLM_M3_types_coefs} suggests that the mobility flows have the strongest positive association with larcenies and vehicle thefts, and a potential negative association with robberies.

\subsection{Robustness checks}


\textbf{Time period.} We test the robustness when varying the period under study: testing the empirical relationship between mobility flows and crime on the 2013 data instead of the 2012 data in San Francisco results in identical findings. All regressors attain similar significance levels across the different specifications, while the best model in terms of AIC is again achieved by PGLM~(1b). Moreover, the sign of the coefficients remains unchanged: check-ins, pass-through flow, and incoming/outgoing flow remain positively linked to crime, while self-loops are negatively associated.


\textbf{Path computation}. Furthermore, we include a robustness analysis, where we base our computation of the shortest paths on the layout of the transportation network, instead of queen's proximity on the network of neighborhoods. For this purpose, we use the San Francisco transit map consisting of all buses, metro trains, streetcars and cable cars routes covering the city.\footnote{\tiny\url{www.sfmta.com/maps/muni-system-map}} This modification does not lead to any changes in our overall results. According to \Cref{table:PGLM_robustness_diff}, pass-trough flow remains positively associated with crime levels at statistically significant levels ($p < 0.001$). Also, in the predictive setup, adding pass-through flow as predictor leads to better results: RF~(1b) outperforms RF~(1a), and RF~(2b) outperforms RF~(2a), both at statistically significant levels ($p < 0.001$) -- see \Cref{table:pred_robustness}. Altogether, this confirms that both our explanatory and predictive results remain robust.


\textbf{Socio-demographic factors.} Finally, we test whether our insights hold in the presence of socio-demographic regressors/predictors. Based on yearly American Community Survey data\footnote{\tiny\url{www.census.gov}.}, we compute metrics for the crime correlates postulated by social disorganization theory \cite{Sampson1997}: concentrated disadvantage, residential stability, and ethnic heterogeneity. In the explanatory approach, their addition has no impact (see \Cref{table:PGLM_robustness_diff}), since their effect was already modeled by the fixed-effect term at census tract level. The $\alpha_i$ term in Eq.~\eqref{eqn:PGLM} already captures the potential census tract level unobserved characteristics that are unlikely to vary over time, such as socio-demographic characteristics. In the predictive approach, their addition slightly improves the prediction error, but does not change the key learnings. As seen in \Cref{table:PGLM_robustness_diff} there are significant improvements of specifications RF~(1b) and RF~(2b) over specifications RF~(1a) and RF~(2a), respectively, i.e., the addition of the path through flow predictor leads to a prediction improvement.

Importantly, a baseline based only on socio-demographic data without the inclusion of any mobility data (but including spatial and temporal features) achieves a MSE = $ 8.068$ on the test set. Hence, it performs poorly and does not match the baseline predictive performance of historical values. Further adding past crime as a feature to this baseline results in a model that achieves a MSE $ = 4.398$ on the test set. Our initial RF~(1a) specification outperforms this baseline by \SI{22.92}{\percent}, while the initial RF~(2b) specification outperforms it by \SI{25.22}{\percent}.

\begin{table}[htbp]
\centering
\footnotesize
{
\begin{tabular}{p{0.3\columnwidth} p{0.2\columnwidth} p{0.2\columnwidth}} 
\toprule
Model   &\multicolumn{2}{c}{Pass-through flow}\\
\cmidrule(rrr){2-3}   
&{PGLM (1b)} &{PGLM (2b)}\\
\midrule
Standard          &\bfseries {7.2\%$^{***}$} &\bfseries {5.1\%$^{***}$}\\
Transportation   &{6.3\%$^{***}$} &{4.5\%$^{***}$}\\
Socio-demographic &{7.2\%$^{***}$} &{5.1\%$^{***}$}\\
\bottomrule
\multicolumn{3}{r}{Significance levels: $^{*}\ 0.05$, $^{**}\ 0.01$, $^{***}\ 0.001$}
\end{tabular}
}
\caption{Robustness check of the explanatory approach based on San Francisco (2012 data).} 
\label{table:PGLM_robustness_diff}
\end{table}

\begin{table}[htbp]
\centering
\footnotesize
{
\begin{tabular}{p{0.3\columnwidth} p{0.12\columnwidth} p{0.12\columnwidth} p{0.12\columnwidth} p{0.12\columnwidth}} 
\toprule
Model   &\multicolumn{2}{c}{MSE} &\multicolumn{2}{c}{MSE}\\
\cmidrule(rrr){2-3}  
\cmidrule(rrr){4-5}  
&{RF (1a)} &{RF (1b)} &{RF (2a)}  &{RF (2b)}\\
\midrule
Standard          &3.390 &3.354  &3.344 &3.289 \\
Transportation    &3.402 &3.296  &3.367 &3.314 \\
Socio-demographic &\bfseries3.370 &\bfseries3.274 &\bfseries3.281 &\bfseries3.190\\
\bottomrule
\end{tabular}
}
\caption{Robustness check of the predictive approach based on San Francisco. The best performance in terms of MSE is highlighted in bold (models are trained on 2012 data and tested on 2013 data).} 
\label{table:pred_robustness}
\end{table}

\section{Discussion}

\subsection{Contributions to theory}


Our results establish that human mobility is highly capable of describing crime concentrations. While check-ins have been studied previously \cite{Kadar2017}, we actually observed an even stronger relationship (\ie, larger coefficient) with pass-through flows. This points to the overall importance of pathways in understanding and theorizing crime. When decomposing check-ins further, we obtained mixed findings: self-loops are negatively associated with crime, whereas the relationship for incoming/outgoing flows is positive. 


The above findings confirm crime pattern theory operationalizing the awareness spaces comprised of both routine activity nodes and pathways via check-ins and pass-through flows, respectively, returns a positive relationship. Accordingly, crime appears to be located at awareness spaces. Even when including an alternative explanation of crime concentrations in cities (i.e., social disorganization theory), our findings remain robust. Finally, we find mixed evidence concerning a natural surveillance according to which humans could function as guardians. This could eventually stir research on testing the eyes-on-the-street theory , but would require additional variables.  

\subsection{Implications} 


This work demonstrates a novel use of data from online location services and it can benefit \textbf{computational social scientists} when investigating urban mobility at a granular, large-scale, and spatio-temporal level. In the context of crime, mining urban mobility information from online location service overcomes the limitations of using traditional survey-based data, which offers only a static view \cite{Felson2015}, and thus entails considerable benefits for theory testing. In fact, it also enables us to decompose different movement patterns such as node activity and pass-through traffic. Our approach makes not use of features that are unique to Foursquare and, therefore, our results should thus generalize to other online services providing location and activity information such as WeChat or Google. 


We have shown how these data from online location services can be utilized towards significantly better predictions of spatio-temporal crime patterns. Provided with access to (near) real-time mobility data, \textbf{public decision-makers} would be able to better respond to short-term evolutions of neighborhood crime. 
For instance, our approach would additionally predict 2,910 crimes per year in San Francisco. With an estimated average dollar loss per burglary of $2,416$ USD\footnote{\tiny\url{ucr.fbi.gov/crime-in-the-u.s/2017/crime-in-the-u.s.-2017/topic-pages/burglary}}, this could amount to savings of up to 7 million USD.

The results from the sensitivity analysis by activity types suggests that people are more exposed during leisure activities. Hence, public and private security forces could choose to improve current security measures in such areas. In comparison, people have lower risk exposure when their awareness space relates to shopping.
Also, drawing from the sensitivity analysis by crime types, policy makers could employ measures for crime prevention targeting especially larcenies and vehicle thefts in less frequented areas.

Furthermore, the implications of this work aid better \textbf{urban designers and policymakers}. In this respect, our work relates to the value of investing in neighborhood cohesion \cite{Sampson1997} and social capital \cite{Hristova2018} in order to bolster neighborhood safety.

\subsection{Limitations and potential for future research} 


It is well established in scientific literature \cite{Arribas-Bel2018} that data from online location services reflects only a subset of the general population - for instance, Foursquare users used to be young, professionals, smartphones owners back in 2012/2013 \cite{Cranshaw2012}. Another potential caveat could be that businesses may have artificially inflated their check-ins.
Today, there may be less users checking in actively on the Foursquare apps, yet, through their SDK-based solution, Foursquare powers location experiences for over 1 billion people on other mobile services. \footnote{\tiny\url{99firms.com/blog/foursquare-statistics/}} Since most check-ins are passive today (\ie, the user is being checked-in as they walk into a place), the number of data points per user is actually higher and the potential for manipulation is lower.

Despite the simplifications we make in regard to mobility features, we can already observe relationships that are statistically significant, even when controlling for time and location fixed-effects. We ran robustness checks across different cities, different types of networks, and also considered alternative predictors. Still, the results might not hold true in smaller urban and rural areas or in other countries.

Our work provides various opportunities for future research. First, our work represents one of the first studies that investigates crime based on hourly resolution; however, even more granular spatio-temporal units could be demanded by practitioners. Second, the work could be advanced by studying potential associations between specific types of crime and urban activities, thereby by propelling further theory building. Third, our results remain to be validated with other sources of granular location data, such as WeChat or Google.

\section{Conclusion}

This work studied the relationship between mobility data from online location services and crime concentrations in novel ways. First, we established that human mobility aids in describing variations in crime concentrations. In fact, we found that both locations of routine activity and pass-through areas reveal a statistically significant influence, while the observed effect size is larger for pathways. Second, our findings contribute to crime theory by advancing our understanding of eyes-on-the-street behavior and finding affirmative evidence of crime pattern theory. Third, we exposed the predictive power of human movements from online location services in relation to crime forecasting. This is likely to propel further innovations in computational social science where routine activities and other social phenomena in urban settings are studied.  
 
\section{Acknowledgements}
We thank Foursquare for supporting this research by providing the dataset used in the analysis.

\bibliography{icwsm}
\bibliographystyle{aaai}

\end{document}